\documentclass[ejs,ps,preprint]{imsart}

\RequirePackage[OT1]{fontenc}
\RequirePackage{amsthm,amsmath}
\RequirePackage[numbers]{natbib}
\RequirePackage[colorlinks,citecolor=blue,urlcolor=blue]{hyperref}

\pubyear{2017}
\volume{0}
\issue{0}
\firstpage{0}
\lastpage{0}

\startlocaldefs
\numberwithin{equation}{section}
\theoremstyle{plain}
\newtheorem{thm}{Theorem}[section]
\newtheorem{cor}{Corollary}[section]
\endlocaldefs

\usepackage{amssymb}
\usepackage{amsmath,amsthm,graphicx,bbm,url}
\usepackage[usenames,dvipsnames]{xcolor}
\usepackage{fancyhdr}
\usepackage{subfig}
\usepackage{enumitem}
\usepackage{float}
\usepackage{subfig}
\usepackage{bm}
\newcommand{\V}[1]{#1} % vector
\newcommand{\M}[1]{#1} % matrix

\usepackage{mathtools}

\newcommand{\Var}{\text{Var}}
\newcommand{\E}{\mathbb{E}}

\newcommand{\argmin}[1]{\underset{#1}{\operatorname{argmin}}\text{ }}

\newcommand{\w}{\alpha}

\newcommand{\ind}[1]{\mathbbm{1}_{#1}}
\newcommand{\rightarrowp}{\overset{P}{\to}}
\newcommand{\rightarrowd}{\overset{d}{\to}}

\newtheorem{Assume}{\underline{\bf Assumptions}}
\def\E{\mathbb{E}}

\begin{document}

\begin{frontmatter}
\title{A Note on Parameter Estimation for Misspecified Regression Models with Heteroskedastic Errors\thanksref{T1}}
\runtitle{Misspecified Regression Models with Heteroskedastic Errors}
\thankstext{T1}{The author thanks the Editor and two reviewers for their constructive
comments.}

\begin{aug}
\author{\fnms{James P.} \snm{Long}\thanksref{t1}\ead[label=e1]{jlong@stat.tamu.edu}}
%% \and
%% \author{\fnms{Second} \snm{Author}\thanksref{t3}\ead[label=e2]{second@somewhere.com}}

%% Department of Statistics, Texas A\&M University\\
%% 3143 TAMU, College Station, TX 77843-3143\\
%% jlong@stat.tamu.edu\\

\address{Department of Statistics\\
%  Texas A\&M University\\
  3143 TAMU\\
  College Station, TX 77843-3143\\
\printead{e1}}

%% \author{\fnms{Third} \snm{Author}
%% \ead[label=e3]{third@somewhere.com}
%% \ead[label=u1,url]{www.foo.com}}

%% \address{Address of the Third author\\
%% usually few lines long\\
%% usually few lines long\\
%% \printead{e3}\\
%% \printead{u1}}

\thankstext{t1}{Long's work was supported by a faculty startup grant from Texas A\&M University.}
%\thankstext{t2}{First supporter of the project}
%\thankstext{t3}{Second supporter of the project}
\runauthor{J.P. Long}

\affiliation{Texas A\&M University}

\end{aug}

\begin{abstract}
Misspecified models often provide useful information about the true data generating distribution. For example, if $y$ is a non--linear function of $x$ the least squares estimator $\widehat{\beta}$ is an estimate of $\beta$, the slope of the best linear approximation to the non--linear function. Motivated by problems in astronomy, we study how to incorporate observation measurement error variances into fitting parameters of misspecified models. Our asymptotic theory focuses on the particular case of linear regression where often weighted least squares procedures are used to account for heteroskedasticity. We find that when the response is a non--linear function of the independent variable, the standard procedure of weighting by the inverse of the observation variances can be counter--productive. In particular, ordinary least squares may have lower asymptotic variance. We construct an adaptive estimator which has lower asymptotic variance than either OLS or standard WLS. We demonstrate our theory in a small simulation and apply these ideas to the problem of estimating the period of a periodic function using a sinusoidal model.
\end{abstract}


\tableofcontents
\end{frontmatter}

\section{Introduction} 
\label{sec:intro}

Misspecified models are common. In prediction problems, simple, misspecified models may be used instead of complex models with many parameters in order to avoid overfitting. In big data problems, true models may be computationally intractable, leading to model simplifications which induce some level of misspecification. In many scientific domains there exist sets of well established models with fast computer implementations. A practitioner with a particular data set may have to choose between using one of these models (even when none are exactly appropriate) and devising, testing and implementing a new model. Pressed for time, the practitioner may use an existing misspecified model. In this work we study how to fit a misspecified linear regression model with heteroskedastic measurement error. Problems involving heteroskedastic measurement error and misspecified models are common in astronomy. We discuss an example in Section \ref{sec:astro}.

Suppose $x_i \in \mathbb{R}^p \sim F_X$ independent across $i$ and $\sigma_i \in \mathbb{R} \sim F_\sigma$ independent across $i$ for $1 \leq i \leq n$. Suppose
\begin{equation*}
y_i = f(x_i) + \sigma_i \epsilon_i
\end{equation*}
where $\epsilon_i \sim F_\epsilon$ with $\E[\epsilon_i] = 0$ and $\Var(\epsilon_i) = 1 \, \forall i$, independent across $i$ and independent of $x_i$ and $\sigma_i$. Define 
\begin{equation*}
\beta \equiv \argmin{\beta} \E[(f(x) - x^T\beta)^2] = \E[xx^T]^{-1}\E[xf(x)].
\end{equation*}
The parameter $\beta$ is the slope of the best fitting least squares line. The parameter $\beta$ may be of interest in several situations. For example, $\beta$ minimizes mean squared error in predicting $y$ from $x$ among all linear functions, ie $\beta = \argmin{\beta} \E[(y - x^T\beta)^2]$. Define $g(x) = f(x) - x^T\beta$. The function $g$ is the non--linear component of $f$.

When the model is correctly specified (ie $g(x) \equiv 0$), weighted least squares (WLS) using the inverse of the observation variances as weights is asymptotically normal and has minimum asymptotic variance among all WLS estimators. In the case with model misspecification and $x_i$, $\sigma_i$ independent, we show that WLS estimators remain asymptotically normal. However weighting by the inverse of the observation variances can result in a larger asymptotic variance than other weightings, including ordinary least squares. Using the asymptotic variance formula we determine an optimal weighting which has lower asymptotic variance than standard WLS (using the inverse of the observation variances as weights) and OLS. The optimal weighting function has the form $w(\sigma) = (\sigma^2 + \Delta)^{-1}$ where $\Delta \geq 0$ is a function of the degree of model misspecification and the design. We find adaptive estimators for $w$ in the cases where the error variances are assumed known and where the error variances belong to one of $M$ groups with group membership known. We also briefly consider the case where $x_i$ and $\sigma_i$ are dependent. In this setting the OLS estimator is consistent but weighted estimators are generally not consistent.

This work is organized as follows. In Section \ref{sec:astro} we introduce a motivating problem from astronomy and offer some heuristic thinking about misspecified models and heteroskedasticity. For those readers primarily interested in the statistical theory, Section \ref{sec:astro} can be skipped. In Section \ref{sec:asymp} we review some relevant literature and develop asymptotic results for the linear model.  We present results for simulated data and the astronomy application in Section \ref{sec:period}. We conclude in Section \ref{sec:discussion}.

\section{Misspecified Models and Heteroskedastic Error in Astronomy}
\label{sec:astro}

\begin{figure}[t]
\centering
 \subfloat[Unfolded Light Curve.]{\label{fig:unfolded}
\includegraphics[scale=.45]{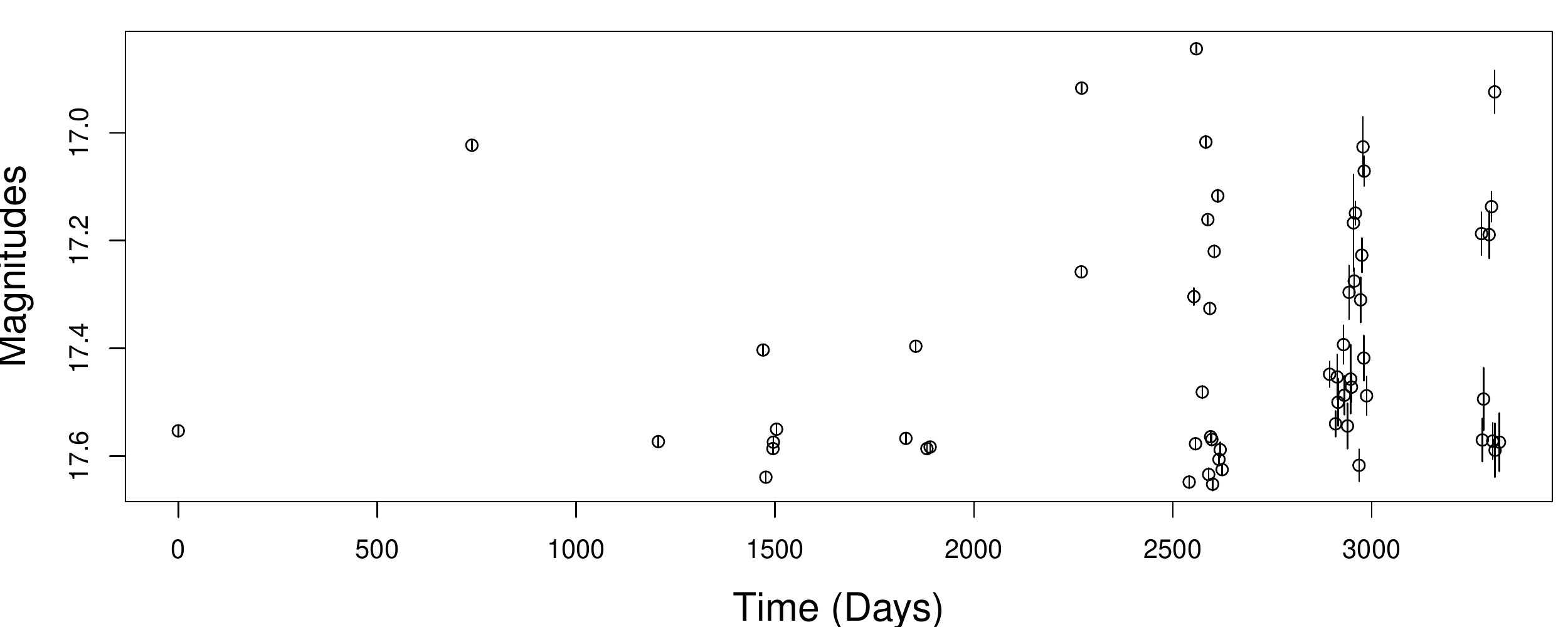}
 } \\ 
 \subfloat[Folded Light Curve.]{\label{fig:folded}
\includegraphics[scale=.45]{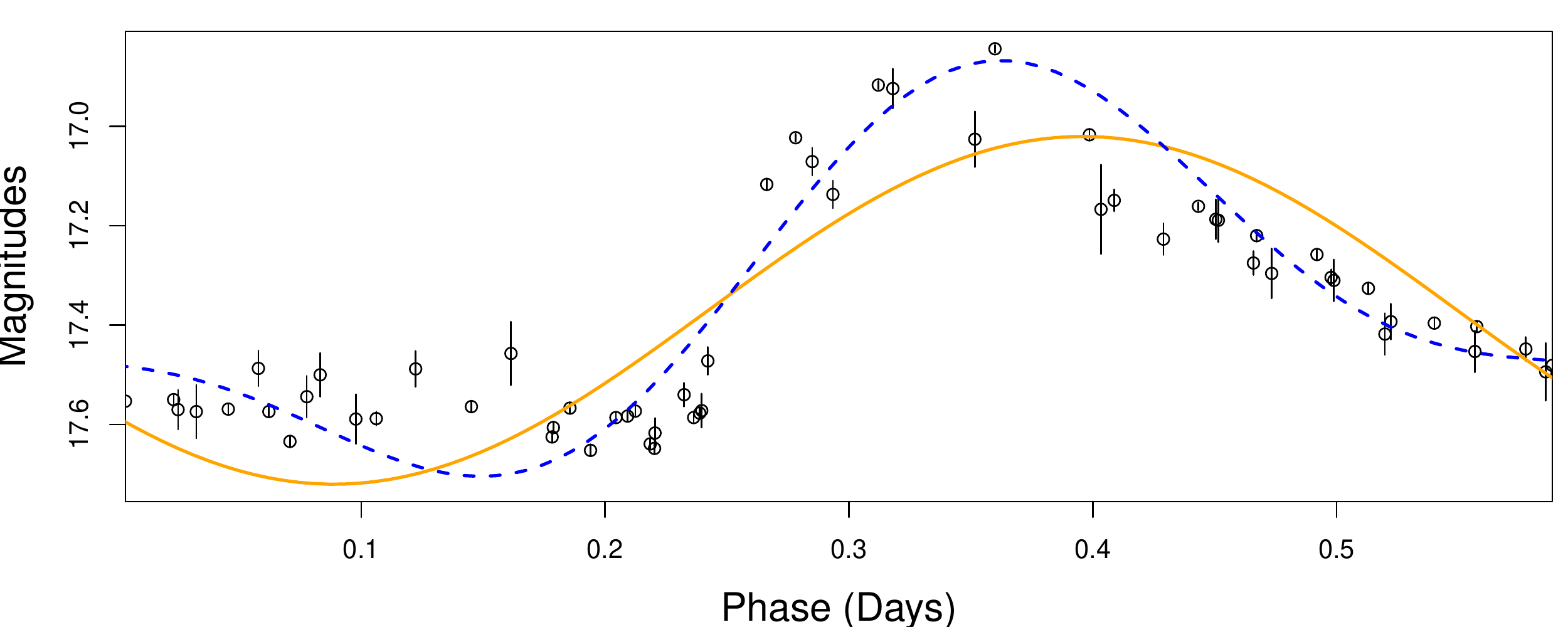}
}
 \caption{(a) SDSS-III RR Lyrae light curve. (b) Folded light curve (x--axis is time modulo period) after estimating the period using the data in a).}
\end{figure}

Periodic variables are stars that vary in brightness periodically over time. Figure \ref{fig:unfolded} shows the brightness of a single periodic variable star over time. This is known as the \textit{light curve} of the star. Two sigma uncertainties are plotted as vertical bars around each point. Magnitude is inversely proportional to brightness, so lower magnitudes are plotted higher on the y--axis. This is a periodic variable so the changes in brightness over time are periodic. Using this data one may estimate a period for the star. When we plot the brightness measurements as time modulo period (Figure \ref{fig:folded}), the pattern in brightness variation becomes clear. Periodic variables play an important role in several areas of astronomy including extra--galactic distance determination and estimation of the Hubble constant \cite{shappee2011new,riess20113}. Modern surveys, such as OGLE-III, have collected hundreds of thousands of periodic variable star light curves \cite{udalski2008optical}.

Accurate period estimation algorithms are necessary for creating the folded light curve (Figure \ref{fig:folded}). A common procedure for determining the period is to perform maximum likelihood estimation using some parametric model for light curve variation. One popular model choice is a sinusoid with $K$ harmonics. Let the data for a single periodic variable be $D = \{(t_{i},y_{i},\sigma_{i})\}_{i=1}^{n}$ where $y_{i}$ is the brightness at time $t_{i}$, measured with known uncertainty $\sigma_{i}$. Magnitude variation is modeled as
\begin{equation}
\label{eq:model}
y_{i} = \beta_{0} + \sum_{k=1}^K a_{k} \sin(k\omega t_{i} + \phi_{k}) + \sigma_i\epsilon_{i}
\end{equation}
where $\epsilon_{i} \sim N(0,1)$ independent across $i$. Here $\omega$ is the frequency, $a_k$ is the amplitude of the $k^{th}$ harmonic, and $\phi_k$ is the phase of the $k^{th}$ harmonic. Let $a=(a_1,\ldots,a_K)$ and $\phi = (\phi_1,\ldots,\phi_K)$. Let $\Omega$ be a grid of possible frequencies. The maximum likelihood estimate for frequency is
\begin{equation}
\label{eq:ml_fit}
\widehat{\omega} = \argmin{\omega \in \Omega} \min_{\V{a},\V{\phi},\beta_0} \sum_{i=1}^{n} \left(\frac{y_{i} - \beta_{0} - \sum_{k=1}^K a_{k} \sin(k\omega t_{i} + \phi_{k})}{\sigma_{i}}\right)^2.
\end{equation}
Generalized Lomb--Scargle (GLS) is equivalent to this estimator with $K=1$ \cite{zechmeister2009generalised}. The analysis of variance periodogram in \cite{schwarzenberg1996fast} uses this model with a fast algorithm for computing $\widehat{\omega}$.

We used estimator \eqref{eq:ml_fit} with $K=1,2$ to determine the period of the light curve in Figure \ref{fig:unfolded}. The estimates for period were essentially the same for both $K=1$ and $K=2$ so in Figure \ref{fig:folded} we folded the light curve using the $K=1$ estimate. The solid orange line is the maximum likelihood fit for the $K=1$ model (notice the sinusoidal shape). The blue dashed line is for the $K=2$ model.

While the period estimates are accurate, both models are misspecified. In particular, note that the vertical lines around the brightness measurements are four standard deviations ($4\sigma_{i}$) in width. If the model is correct, we would expect about 95\% of these intervals to contain the maximum likelihood fitted curves. For the $K=1$ model, 10\% of the intervals contain the fitted curve. For $K=2$, 37\% of observations contain the ML fitted curve. The source of model misspecification is the light curve shape which cannot be perfectly represented by a sinusoid with $K=1,2$ harmonics. The light curve has a long, slow decline and a sudden, sharp increase in brightness.

The parameter fits of misspecified models are estimates of an approximation. In the $K=1$ case, the parameter fits are the orange line in Figure \ref{fig:folded} and the approximation is the sinusoid which is closest to the true light curve shape. In many cases this approximation may be useful. For example the period of the approximation may match the period of the light curve.

When fitting a misspecified model with heteroskedastic measurement error, one should choose a weighting which ensures the estimator has small variance and thus is likely close to the approximation. The use of the inverse of the observation variances as weights (in Equation \eqref{eq:beta_est}) is motivated by maximum likelihood theory under the assumption that the model is correct. However as we show in Section \ref{sec:asymp} for the linear model, these weights are generally not optimal when there is model misspecification.

As a thought experiment, consider the case where one observation has extremely small variance and other observations have much larger variance. The maximum likelihood fitted curve for this data will be very close to the observation with small variance. However the best sinusoidal approximation to the true function at this point may not be particularly close to the true function. Thus using the inverse of observation variances as weights may overweight observations with small variance in the case of model misspecification. We make these ideas precise in Section \ref{sec:olswls}.

The choice of weights is not critical for the light curve in Figure \ref{fig:unfolded} because it is well sampled ($n > 50$), so the period is easy to determine. However in many other cases light curves are more poorly sampled ($n \approx 20$), in which case weighting may affect period estimation accuracy.

\subsection{Sinusoidal Fit and Linear Models}
Finding the best fitting sinusoid is closely related to fitting a linear model. Using the sine angle addition formula we can rewrite the maximum likelihood estimator from Equation \eqref{eq:ml_fit} as
\begin{equation*}
\argmin{\omega \in \Omega} \min_{\V{a},\V{\phi},\beta_0} \sum_{i=1}^{n} \left(\frac{y_{i} - \sum_{k=1}^K (a_{k}\cos(\phi_{k})\sin(k\omega t_{i})  + a_{k}\sin(\phi_{k})\cos(k\omega t_{i})) - \beta_{0}}{\sigma_{i}}\right)^2
\end{equation*}
The sum over $i$ can be simplified by noting the linearity of the model and reparameterizing. Let $Y = (y_{1},\ldots,y_{n})^T$. Let $\beta_{k1} = a_{k}\cos(\phi_{k})$ and $\beta_{k2} = a_{k}\sin(\phi_{k})$. Define $\beta = (\beta_{0},\beta_{11},\beta_{12},\ldots,\beta_{K1},\beta_{K2})^T \in \mathbb{R}^{2K+1}$. Let $\M{\Sigma}$ be a $n \times n$ diagonal matrix where $\Sigma_{ii} = \sigma_{i}^2$. Define
\begin{equation*}
\M{X}(\omega) = \begin{pmatrix}
 1 & \sin(\omega t_{1}) & \cos(\omega t_{1}) & \dots  & \sin(K\omega t_{1}) & \cos(K\omega t_{1}) \\
 1 & \sin(\omega t_{2}) & \cos(\omega t_{2}) & \dots  & \sin(K\omega t_{2}) & \cos(K\omega t_{2}) \\
 \vdots  & \vdots  &\vdots  &\ddots & \vdots & \vdots  \\
 1 & \sin(\omega t_{n}) & \cos(\omega t_{n}) & \dots  & \sin(K\omega t_{n}) & \cos(K\omega t_{n}) \\
\end{pmatrix} \in \mathbb{R}^{n \times (2K+1)}.
\end{equation*}
We rewrite the ML estimator as
\begin{equation*}
\widehat{\omega} = \argmin{\omega \in \Omega} \min_{\V{\beta}} \, (\V{Y} - \M{X}(\omega)\V{\beta})^T\M{\Sigma}^{-1}(\V{Y} - \M{X}(\omega)\V{\beta})
\end{equation*}
Every frequency $\omega$ in the grid of frequencies $\Omega$ determines a design matrix $\M{X}(\omega)$. At a particular $\omega$, the $\beta$ which minimizes the objective function is the weighted least squares estimator
\begin{equation}
\label{eq:beta_est}
\widehat{\V{\beta}}(\omega) = (\M{X}(\omega)\M{\Sigma}^{-1}\M{X}(\omega))^{-1}\M{X}(\omega)^T\M{\Sigma}^{-1}Y.
\end{equation}
The frequency estimate may then be written as
\begin{equation}
\label{eq:freq_est}
\widehat{\omega} = \argmin{\omega \in \Omega}  \, (\V{Y} - \M{X}(\omega)\widehat{\V{\beta}}(\omega))^T\M{\Sigma}^{-1}(\V{Y} - \M{X}(\omega)\widehat{\V{\beta}}(\omega)).
\end{equation}

Thus estimating frequency involves performing a weighted least squares regression (Equation \eqref{eq:beta_est}) at every frequency in the grid $\Omega$. The motivation for the procedure is maximum likelihood. As discussed earlier, in cases where the model is misspecified, there is no theoretical support for using $\M{\Sigma}^{-1}$ as the weight matrix in either Equation \eqref{eq:beta_est} or \eqref{eq:freq_est}.

\section{Asymptotic Theory}
\label{sec:asymp}

\subsection{Problem Setup and Related Literature}

Let $X \in \mathbb{R}^{n \times p}$ be the matrix with row $i$ equal to $x_i^T$. Let $Y=(y_1,\ldots,y_n)^T$. Let $\Sigma$ be the diagonal matrix of observation variances such that $\Sigma_{ii} = \sigma_i^2$. Let $\widehat{W}$ be a diagonal positive definite matrix. The weighted least squares estimator is
\begin{equation*}
\widehat{\beta}(\widehat{W}) = (X^T\widehat{W}X)^{-1}X^T\widehat{W}Y.
\end{equation*}
In this work we seek $\widehat{W}$ which minimize error in estimating $\beta = \E[xx^T]^{-1}\E[xf(x)]$.

There is a long history of studying estimators for misspecified models, often in the context of sandwich estimators for asymptotic variances. In \cite{huber1967behavior}, it was shown that when the true data generating distribution $\theta_t$ is not in the model, the MLE converges to the distribution $\theta_0$ in the model $\Theta$ which minimizes Kullback--Liebler divergence, ie
\begin{equation*}
\theta_0 = \argmin{\theta \in \Theta} \E_{\theta_t}\left[\frac{\log f_{\theta_t}(X)}{\log f_\theta(X)}\right].
\end{equation*}
The asymptotic variance has a ``sandwich'' form which is not the inverse of the information matrix. \cite{white1980heteroskedasticity} and \cite{white1980using} studied this behavior in the context of the linear regression model and the OLS estimator, proposing consistent estimators of the asymptotic variance. See \cite{mackinnon1985some} and \cite{long2000using} for sandwich estimators with improved finite sample performance and \cite{szpiro2010model} for recent work on sandwich estimators in a Bayesian context. \cite{buja2014models} provides a summary of sandwich estimators and proposes a bootstrap estimator for the asymptotic variance. By specializing our asymptotic theory from the weighted to the unweighted case, we rederive some of these results. However our focus is different in that we find weightings for least squares which minimize asymptotic variance, rather than estimating the asymptotic variance of unweighted procedures.

Other work has focused on correcting model misspecification, often by modeling deviations from a parametric regression function with some non--parametric model. \cite{blight1975bayesian} studied model misspecification when response variances are known up to a constant due to repeated measurements, ie $Var(y_i) = \sigma^2/m_i$ where $m_i$ is known. A Gaussian prior was placed on $\beta$ and the non--linear component $g$ was modeled as being drawn from a Gaussian process. See \cite{kennedy2001bayesian} for an example with homoskedastic errors in the context of computer simulations. See \cite{czekala2015constructing} for an example in astronomy with known heteroskedastic errors. Our focus here is different in that instead of correcting model misspecification we consider how weighting observations affects estimation of the linear component of $f$. 

Heteroskedasticity in the partial linear model
\begin{equation*}
y_i = x_i^T\beta + h(z_i) + \epsilon_i.
\end{equation*}
is studied in \cite{ma2006efficient} and \cite{ma2013doubly}. Here $\Var(\epsilon_i) = \xi(x_i,z_i)$ for some function $\xi$. The parameter $h$ is some unknown function. The response $y$ depends on the $x$ covariates linearly and the $z$ covariates nonlinearly. When $h$ is estimated poorly, weighting by the inverse of the observation variances causes parameter estimates of $\beta$ to be inconsistent. In contrast, ignoring observation variances leads to consistent estimates of $\beta$.  Qualitatively these conclusion are similar to our own in that they caution against using weights in the standard way.

\subsection{Asymptotic Results}

Our asymptotic theory makes assumptions on the form of the weight matrix.
\begin{Assume}[Weight Matrix]
\label{ass:weight}
Suppose $\widehat{W} \in \mathbb{R}^n \times \mathbb{R}^n$ is a positive definite diagonal matrix with elements
\begin{equation*}
\widehat{W}_{ii} = w(\sigma_i) + n^{-1/2}\delta_{nm_i}h(\sigma_i) + n^{-1}d(\sigma_i,\delta_n)
\end{equation*}
where $w(\sigma) > 0$, $\E[w(\sigma)^4] < \infty$, $h$ is a bounded function, $m_i \in \{1,\ldots,M\}$ is a discrete random variable independent of $x_i$ and $\epsilon_i$, $\delta_{nm_i}$ is $O_P(1)$ for all $n \in \mathbb{Z}^+$ and $m_i$, and $d(\sigma,\delta_n)$ is uniformly in $\sigma$ bounded above by an $O_P(1)$ random variable (ie $\sup_{\sigma} |d(\sigma,\delta_n)| < \delta_n'$ where $\delta_n'$ is $O_P(1)$).
\end{Assume}

These assumptions include both the ordinary least squares (OLS) estimator where $\widehat{W}_{ii} = w(\sigma_i) = 1$ and the standard weighted least squares estimator where $\widehat{W}_{ii} = w(\sigma_i) = \sigma_i^{-2}$ (assuming $\E[\sigma^{-8}] < \infty$). In both these cases $\delta_{nm_i} = 0$ and $d = 0$ for all $n,m$. These additional terms are used in Sections \ref{sec:known} and \ref{sec:unknown} to construct adaptive estimators for the known and unknown variance cases.

\begin{Assume}[Moment Conditions]
\label{ass:moment}
Suppose $x$ and $\sigma$ are independent, the design $\E[xx^T]$ is full rank, and $\E[x_j^4x_k^4] < \infty$ for all $1 \leq j,k \leq p$. Assume $\E[g(x)^4] < \infty$, $\E[\sigma^4] < \infty$, $\E[\epsilon^4] < \infty$, and the variances are bounded below by a positive constant $\sigma_{min}^2 \equiv \inf \{\sigma^2 : F_\sigma(\sigma) > 0\} > 0$.
\end{Assume}
The major assumption here is independence between $x$ and $\sigma$. We address dependence in Section \ref{sec:dependent}.
\begin{thm}
\label{thm:clt}
Under Assumptions \ref{ass:weight} and \ref{ass:moment}
\begin{equation*}
\sqrt{n}(\widehat{\beta}(\widehat{W}) - \beta) \rightarrowd N(0,\nu(w))
\end{equation*}
where
\begin{equation}
\label{eq:asymp_var}
\nu(w) = \frac{\E[w^2]\E[xx^T]^{-1}\E[g^2(x)xx^T]\E[xx^T]^{-1} + \E[\sigma^2w^2]\E[xx^T]^{-1}}{\E[w]^2}.
\end{equation}
\end{thm}
See Section \ref{prf:clt} for a proof. If the response is linear ($g \equiv 0$) then the variance is
\begin{equation*}
\nu(w) = \frac{\E[\sigma^2w^2]}{\E[w]^{2}}\E[xx^T]^{-1}.
\end{equation*}
Setting $w(\sigma) = \sigma^{-2}$ we have $\frac{\E[\sigma^2w^2]}{\E[w]^{2}} = (\E[\sigma^{-2}])^{-1}$. This is the standard weighted least squares estimator. This $w$ can be shown to minimize the variance using the Cauchy Schwartz inequality. With $w(\sigma)=1$, the asymptotic variance can be rewritten
\begin{equation}
  \label{eq:sandwich}
\E[xx^T]^{-1}\E[(g^2(x)+\sigma^2)xx^T]\E[xx^T]^{-1}.
\end{equation}
This is the sandwich form of the covariance for OLS derived in \cite{white1980heteroskedasticity} and \cite{white1980using} (see \cite{buja2014models}, specifically Equations 1-3), valid even when $\sigma$ and $x$ are not independent.

\subsection{OLS and Standard WLS}
\label{sec:olswls}
For notational simplicity define
\begin{align*}
B &= \E[xx^T]^{-1}\\
A &= B^T\E[g^2(x)xx^T]B.
\end{align*}
The asymptotic variances for OLS ($\widehat{\beta}(I)$) and standard WLS ($\widehat{\beta}(\Sigma^{-1})$) are
\begin{align*}
\nu(I) &= A + \E[\sigma^2]B\\
\nu(\Sigma^{-1}) &= \frac{\E[\sigma^{-4}]}{\E[\sigma^{-2}]^2}A + \frac{1}{\E[\sigma^{-2}]}B.
\end{align*}
Each of these asymptotic variances is composed of the same two terms. The $A$ term is caused by model misspecification while the $B$ term is the standard asymptotic variance in the case of no model misspecification. The coefficient on $A$ is larger for $W=\Sigma^{-1}$ because $\frac{\E[\sigma^{-4}]}{\E[\sigma^{-2}]^2} \geq 1$ by Jensen's Inequality. The coefficient on $B$ is larger for $W=I$ because $\E[\sigma^2] \geq \frac{1}{\E[\sigma^{-2}]}$. The relative merits of OLS and standard WLS depend on the size of the coefficients and the precise values of $A$ and $B$. However, qualitatively, OLS and standard WLS suffer from high asymptotic variance in opposite situations which depend on the distribution of the errors. To make matters concrete, consider error distributions of the form
\begin{align*}
&P(\sigma=c^{-1}) = \delta_1\\
&P(\sigma=1) = 1 - \delta_1 - \delta_2\\
&P(\sigma=c) = \delta_2
\end{align*}
where $\delta_1,\delta_2$ are small non--negative numbers and $c > 1$ is large. Note that $A$ and $B$ do not depend on $F_\sigma$. 
\begin{itemize}
\item \textbf{$\delta_1=0, \delta_2 > 0$:} In this situation the error standard deviation is usually $1$ and occasionally some large value $c$. The result is large asymptotic variance for OLS. Since $\E[\sigma^2] > c^2\delta_2$, 
\begin{equation*}
\nu(I) \succeq A + c^2\delta_2B
\end{equation*}
For large $c$ this will be large. In contrast the coefficients on $A$ and $B$ for standard WLS can be bounded. For the coefficient on $B$ we have $\E[\sigma^{-2}]^{-1} \leq (1-\delta_2)^{-1}$. The coefficient on $A$ with $c > 1$ is
\begin{equation*}
\frac{\E[\sigma^{-4}]}{\E[\sigma^{-2}]^2} = \frac{\delta_2c^{-4} + (1-\delta_2)}{\delta_2^2c^{-4} + 2\delta_2c^{-2}(1-\delta_2) + (1-\delta_2)^2} < \frac{1}{1-\delta_2}.
\end{equation*}
Therefore
\begin{equation*}
\nu(\Sigma^{-1}) \preceq (1-\delta_2)^{-1}(A + B). 
\end{equation*}
In summary, standard WLS performs better than OLS when there are a small number of observations with large variance.
\item \textbf{$\delta_1>0,\delta_2=0$:} In this situation the error standard deviation is usually $1$ and occasionally some small value $c^{-1}$. For standard WLS with $c$ large and $\delta_1$ small, the coefficient for $A$ is
\begin{equation*}
\frac{\E[\sigma^{-4}]}{\E[\sigma^{-2}]^2} = \frac{\delta_1c^4 + (1-\delta_1)}{\delta_1^2c^4 + 2\delta_1c^2(1-\delta_1) + (1-\delta_1)^2} \approx \frac{1}{\delta_1}.
\end{equation*}
Thus the asymptotic variance induced by model misspecification will be large for standard WLS. In contrast, we can bound the asymptotic variance above for OLS, independently of $c$ and $\delta_1$. Since $c > 1$, $\E[\sigma^2] < 1$ and 
\begin{equation*}
\nu(I) \preceq A + B.
\end{equation*}
\end{itemize}
The case where both $\delta_1$ and $\delta_2$ are non--zero presents problems for both OLS and standard WLS. For example if $\delta = \delta_1 = \delta_2$, both OLS and standard WLS can be made to have large asymptotic variance by setting $\delta$ small and $c$ large. In the following section we construct an adaptive weighting which improves upon both OLS and standard WLS.

\subsection{Improving on OLS and Standard WLS}
\label{sec:improve}
Let $\Gamma$ be a linear function from the set of $p \times p$ matrices to $\mathbb{R}$ such that $\Gamma(C) > 0$ whenever $C$ is positive definite. We seek some weighting $w = w(\sigma)$ for which $\Gamma(\nu(w))$ (recall that $\nu$ is the asymptotic variance) is lower than OLS and standard WLS. Natural choices for $\Gamma$ include the trace (minimize the sum of variances of the parameter estimates) and the $\Gamma(C) = C_{jj}$ (minimize the variance of one of the parameter estimates).

\begin{thm}
\label{thm:lagrange}
Under Assumptions \ref{ass:weight} and \ref{ass:moment}, every function in the set
\begin{equation*}
\argmin{w(\sigma)} \Gamma(\nu(w))
\end{equation*}
is proportional to
\begin{equation}
\label{eq:wmin}
w_{min}(\sigma) = (\sigma^2 + \Gamma(A)\Gamma(B)^{-1})^{-1}
\end{equation}
with probability $1$.
\end{thm}
Section \ref{prf:lagrange} contains a proof. The proportionality is due to the fact that the estimator is invariant to multiplicative scaling of the weights.

\begin{cor}
\label{thm:adaptive}
Under Assumptions \ref{ass:moment},
\begin{equation*}
\Gamma(\nu(w_{min})) \leq \min(\Gamma(\nu(I)),\Gamma(\nu(\Sigma^{-1})))
\end{equation*}
with strict inequality if $\E[g^2(x)xx^T]$ is positive definite and the distribution of $\sigma$ is not a point mass.
\end{cor}
A proof is contained in Section \ref{prf:adaptive}. Thus if we can construct a weight matrix $\widehat{W}$ which satisfies Assumptions \ref{ass:weight} with $w(\sigma) = w_{min}(\sigma)$ , then by the preceding theorem the associated weighted estimator will have lower asymptotic variance then either OLS or standard WLS. We now construct such a weighting in the case of known and unknown error variances.

\subsection{Known Error Variances}
\label{sec:known}

With the $\sigma_i$ known we only need to estimate $A$ and $B$ in $w_{min}$ in Equation \eqref{eq:wmin}. Let $\Delta = \Gamma(A)\Gamma(B)^{-1}$. Let
\begin{equation*}
\widehat{B} = \left(\frac{1}{n}X^TX\right)^{-1}.
\end{equation*}
Let $\widehat{\beta}(\widehat{W})$ be a root $n$ consistent estimator of $\beta$ (eg $\widehat{W}=I$ is root $n$ consistent by Theorem \ref{thm:clt}) and let
\begin{equation*}
\widehat{g}(x_i)^2 = (y_i - x_i^T\widehat{\beta}(\widehat{W}))^2 - \sigma_i^2.
\end{equation*}
Let
\begin{equation*}
\widehat{A} = \widehat{B}^T\left(\sum \sigma_i^{-4}\right)^{-1}\left(\sum x_ix_i^T\widehat{g}(x_i)^2\sigma_i^{-4}\right)\widehat{B}.
\end{equation*}
Then we have
\begin{equation}
\label{eq:deltest}
\widehat{\Delta} = \max(\Gamma(\widehat{A})\Gamma(\widehat{B})^{-1},0).
\end{equation}
The estimated optimal weighting matrix is the diagonal matrix $\widehat{W}_{min}$ with diagonal elements
\begin{equation}
\label{eq:est_opt}
\widehat{W}_{min,ii} = \frac{1}{\sigma_i^2 + \widehat{\Delta}}.
\end{equation}
A few notes on this estimator:
\begin{itemize}
\item The term $x_ix_i^T\widehat{g}(x_i)^2$ is an estimate of $x_ix_i^Tg(x_i)^2$. These estimates are weighted by $\sigma_i^{-4}$. The term $(\sum \sigma_i^{-4})^{-1}$ normalizes the weights. This weighting is motivated by the fact that
\begin{align*}
x_ix_i^T\widehat{g}(x_i)^2 &= x_ix_i^T((y_i - x_i^T\widehat{\beta}(\widehat{W}))^2 - \sigma_i^2) \\
&= x_ix_i^T((y_i - x_i^T\beta)^2 - \sigma_i^2) + O(n^{-1/2})\\
& = x_ix_i^T((g(x_i) + \sigma_i\epsilon_i)^2 - \sigma_i^2) + O(n^{-1/2}).
\end{align*}
Analysis of the first order term shows
\begin{equation*}
  \E[x_ix_i^T((g(x_i) + \sigma_i\epsilon_i)^2 - \sigma_i^2)|x_i,\sigma_i] =  x_ix_i^Tg^2(x_i)
\end{equation*}
and
\begin{align*}
  &\Var(x_ix_i^T((g(x_i) + \sigma_i\epsilon_i)^2 - \sigma_i^2)|x_i,\sigma_i)_{jk} \\
  &=  x_{ij}^2x_{ik}^2(\sigma_i^4(\E[\epsilon^4] - 1) + 4g(x_i)^2\sigma_i^2 + 4g(x_i)\sigma_i^3\E[\epsilon^3])
\end{align*}
Thus by weighting the estimates by $\sigma_i^{-4}$, we can somewhat account for the different variances. Unfortunately since the variance depends on $g$, $\E[\epsilon^3]$, and $\E[\epsilon^4]$ which are unknown, it is not possible to weight by exactly the inverse of the variances. Other weightings are possible and in general adaptivity will hold. 
\item Since $A$ and $B$ are positive semi--definite, $\Gamma(A)\Gamma(B)^{-1} \geq 0$. Thus for estimating $\Delta$, we use the maximum of a plug--in estimator and $0$ (Equation \eqref{eq:deltest}).
\end{itemize}

\begin{thm}
\label{thm:satisfy}
Under Assumptions \ref{ass:moment}, $\widehat{W}_{min}$ from Equation \eqref{eq:est_opt} satisfies Assumptions \ref{ass:weight} with $w(\sigma)=w_{min}(\sigma)$.
\end{thm}
See Section \ref{prf:satisfy} for a proof. Theorem \ref{thm:satisfy} shows it is possible to construct better estimators than both OLS and standard WLS. In practice, it may be best to iteratively update estimates of $\widehat{W}_{min}$ starting with a known root $n$ consistent estimator such as $\widehat{W} = I$. We take this approach in our numerical simulations in Section \ref{sec:sim}.

For the purposes of making confidence regions we need estimators of the asymptotic variance. Above we developed consistent estimators for $A$ and $B$. We take a plug--in approach to estimating the asymptotic variance for a particular weighting $W$. Specifically
\begin{equation}
  \label{eq:nuhat1}
\widehat{\nu}_1(\widehat{W}) = \frac{n(1^T\widehat{W}^21)\widehat{A} + n(1^T\widehat{W}\Sigma\widehat{W}1)\widehat{B}}{(1^T\widehat{W}1)^2}.
\end{equation}
We also define the oracle $\widehat{\nu}_{OR}(\widehat{W})$ which is the same as $\widehat{\nu}_1$ but uses $A$ and $B$ rather than $\widehat{A}$ and $\widehat{B}$. While $\widehat{\nu}_{OR}$ cannot be used in practice, it is useful for evaluating the performance of $\widehat{\nu}_1$ in simulations.

Finally suppose the error variance is known up to a constant, i.e. $\sigma_i^2 = k \tau_i^2$ where $\tau_i^2$ is known but $k$ and $\sigma_i^2$ are unknown. In the case without model misspecification, one can simply use weights $\tau_i^{-2}$ since the weighted estimator is invariant up to rescaling of the weights. The situation is more complicated when model misspecification is present. Simulations and informal mathematical derivations (not included in this work) suggest that replacing the $\sigma_i$ with $\tau_i$ in Equation \eqref{eq:est_opt} results in weights that are suboptimal. In particular, when $k > 1$ (underestimated errors), the resulting weights are closer to OLS than optimal while if $k < 1$ (overestimated errors), the resulting weights are closer to standard WLS than optimal.

\subsection{Unknown Error Variances}
\label{sec:unknown}

Suppose for observation $i$ we observe $m_i \in \{1,\ldots,M\}$, the group membership of observation $i$. Observations in group $m$ have the same (unknown) variance $\sigma_{m}^2 > 0$. See \cite{fuller1978estimation}, \cite{chen1993iterative}, and \cite{hooper1993iterative} for work on grouped error models in the case where the response is linear.

The $m_i$ are assumed independent of $x_i$ and $\epsilon_i$, with probability mass function $f_m$ (supported on $1,\ldots,M$). While the $\sigma_m$ for $m=1,\ldots,M$ are fixed unknown parameters, the probability mass function $f_m$ induces the probability distribution function $F_\sigma$ on $\sigma$. So we can define
\begin{equation*}
\E[h(\sigma)] = \sum_{m=1}^M h(\sigma_m) f_m(m)
\end{equation*}
for any function $h$.

Theorem \ref{thm:clt} shows that even if the $\sigma_{m}$ were known, standard weighted least squares is not generally optimal for estimating $\beta$ in this model. It is not possible to estimate $w_{min}$ as proposed in Section \ref{sec:known} because that method requires knowledge of $\sigma_m$. However we can re--express the optimal weight function as
\begin{align*}
w_{min}(m) &= \frac{1}{\sigma_m^2 + \frac{\Gamma(B^T\E[g^2(x)xx^T]B)}{\Gamma(B)}}\\
&= \frac{\Gamma(B)}{\Gamma(B^T\E[(g^2(x)+\sigma_m^2)xx^T]B)}\\
&= \frac{\Gamma(B)}{\Gamma(B^TC_mB)}
\end{align*}
where the last equality defines $C_m$. Note that $\sigma_m$ is a fixed unknown parameter, not a random variable. One can estimate $B$ with $\widehat{B} = (n^{-1}X^TX)^{-1}$ and $C_m$ with
\begin{equation*}
\widehat{C}_m = \frac{1}{\sum_{i=1}^n \ind{m_i=m}} \sum_{i=1}^n (y_i - x_i^T\widehat{\beta}(\widehat{W}))^2x_ix_i^T\ind{m_i=m}
\end{equation*}
where $\widehat{\beta}(\widehat{W})$ is a root $n$ consistent estimator of $\beta$ (for example $\widehat{W} = I$ suffices by Theorem \ref{thm:clt}). The estimated weight matrix $\widehat{W}_{min}$ is diagonal with 
\begin{equation}
\label{eq:estimated_optimal}
\widehat{W}_{min,ii} =  \frac{\Gamma(\widehat{B})}{\Gamma(\widehat{B}^T\widehat{C}_{m_i}\widehat{B})}.
\end{equation}
\begin{thm}
\label{thm:satisfy2}
Under Assumptions \ref{ass:moment}, $\widehat{W}_{min}$ from Equation \eqref{eq:estimated_optimal} satisfies Assumptions \ref{ass:weight} with $w(\sigma) = w_{min}(m)$.
\end{thm}
See Section \ref{prf:satisfy2} for a proof. Thus in the case of unknown errors it is possible to construct an estimator which outperforms standard WLS and OLS. As is the case with known errors, one can iteratively update $\widehat{W}_{min}$, starting with some (possibly inefficient) root $n$ consistent estimate of $\beta$.

For estimating the asymptotic variance we cannot use Equation \eqref{eq:nuhat1} because that method required an estimate of $A$, a quantity for which we do not have an estimate in the unknown error variance setting. Instead note that the asymptotic variance of Equation \eqref{eq:asymp_var} may be rewritten
\begin{equation*}
  \nu(W) = \frac{B \E[(g^2(x) + \sigma^2)w^2xx^T] B}{\E[w]^2} = \frac{B \E[(y - x^T\beta)^2w^2xx^T] B}{\E[w]^2}.
\end{equation*}
Thus a natural estimator for the asymptotic variance is
\begin{equation}
  \label{eq:nuhat2}
\widehat{\nu}_2(\widehat{W}) = \frac{n\widehat{B} \left(\sum_{i=1}^n (y_i - x_i^T\widehat{\beta}(\widehat{W}))^2\widehat{W}_{ii}^2x_ix_i^T\right) \widehat{B}}{(1^T\widehat{W}1)^2}.
\end{equation}

\subsection{Dependent Errors}
\label{sec:dependent}

Suppose one drops the independence assumption between $x$ and $\sigma$. This will be the case whenever the error variance is a function of $x$, a common assumption in the heteroskedasticity literature \cite{carroll1982adapting,carroll1982robust,jobson1980least}. We require the weight matrix $W$ to be diagonal positive definite with diagonal elements $W_{ii} = w(\sigma_i)$, some function of the error variance. The estimator for $\beta$ is 
\begin{equation*}
\widehat{\beta}(W) = (X^TWX)^{-1}X^TWY.
\end{equation*}
Recalling we write $w$ for $w(\sigma)$, we have the following result.
\begin{thm}
\label{thm:dep}
Assuming $\E[xx^Tw]$, $\E[wxf(x)]$, and $\E[xw\sigma]$ exist and $\E[xx^T]$ is positive definite,
\begin{equation}
  \label{eq:dep_errors}
\widehat{\beta}(W) \rightarrow_{a.s.} \E[xx^T w]^{-1}\E[wxf(x)].
\end{equation}
\end{thm}
See Section \ref{prf:dep} for a proof. If $x$ and $\sigma$ are independent then the r.h.s is $\E[xx^T]^{-1}\E[xf(x)]$ and the estimator is consistent (as demonstrated by Theorem \ref{thm:clt}). Interestingly the estimator is also consistent if one lets $w(\sigma) = 1$ (OLS), regardless of the dependence structure between $x$ and $\sigma$. However weighted estimators will not generally be consistent (including standard WLS). This observation suggests the OLS estimator may be preferred in the case of dependent errors. We show an example of this situation in the simulations of Section \ref{sec:sim}.

\section{Numerical Experiments}
\label{sec:period}

\subsection{Simulation}
\label{sec:sim}

\def \w {.35}
  \begin{figure}[H]
\centering
 \subfloat[$\widehat{W}=\Sigma^{-1}$]{\label{fig:estimator_wls}
\includegraphics[scale=\w]{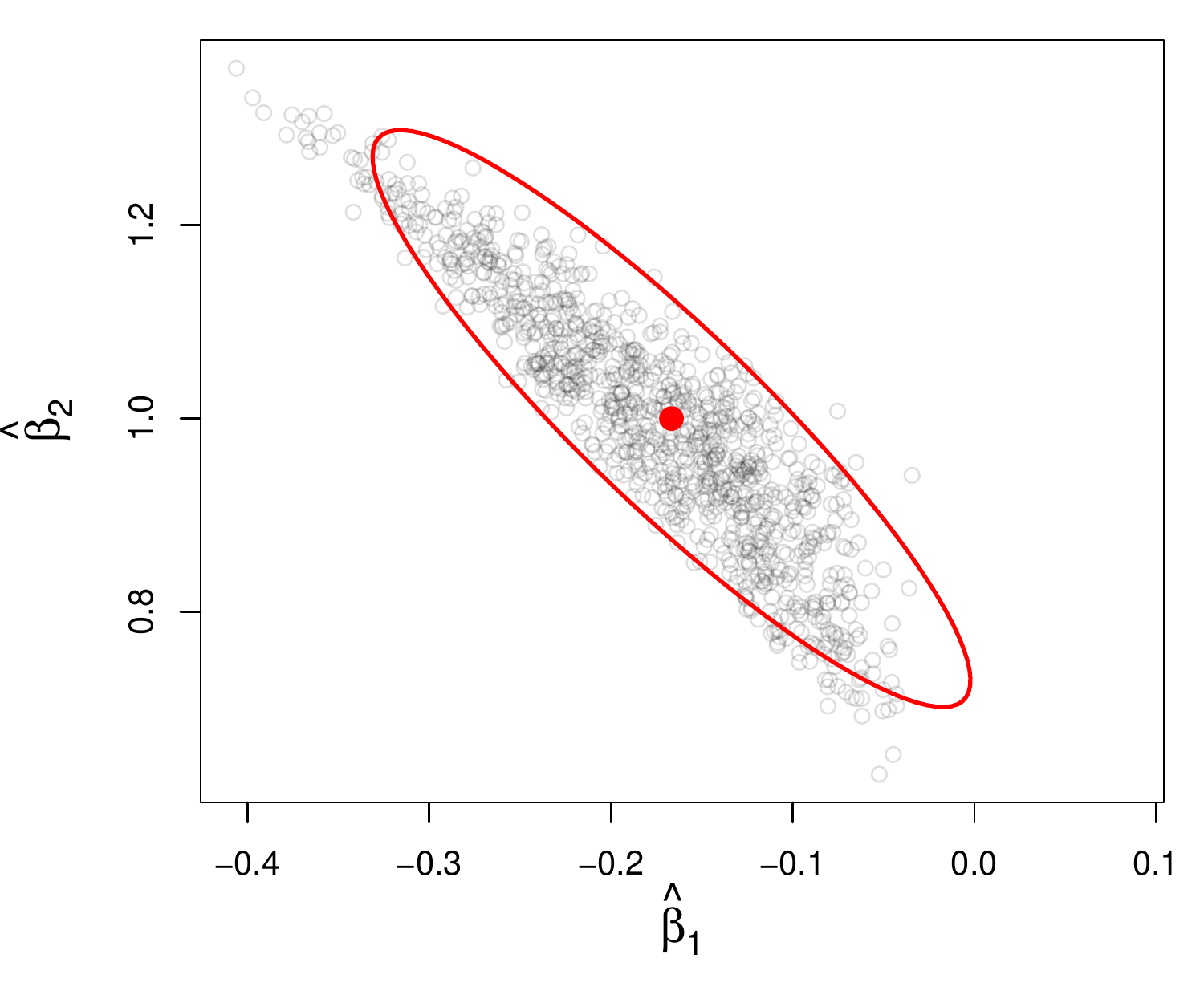}
}
\subfloat[$\widehat{W}=I$]{\label{fig:estimator_ols}
\includegraphics[scale=\w]{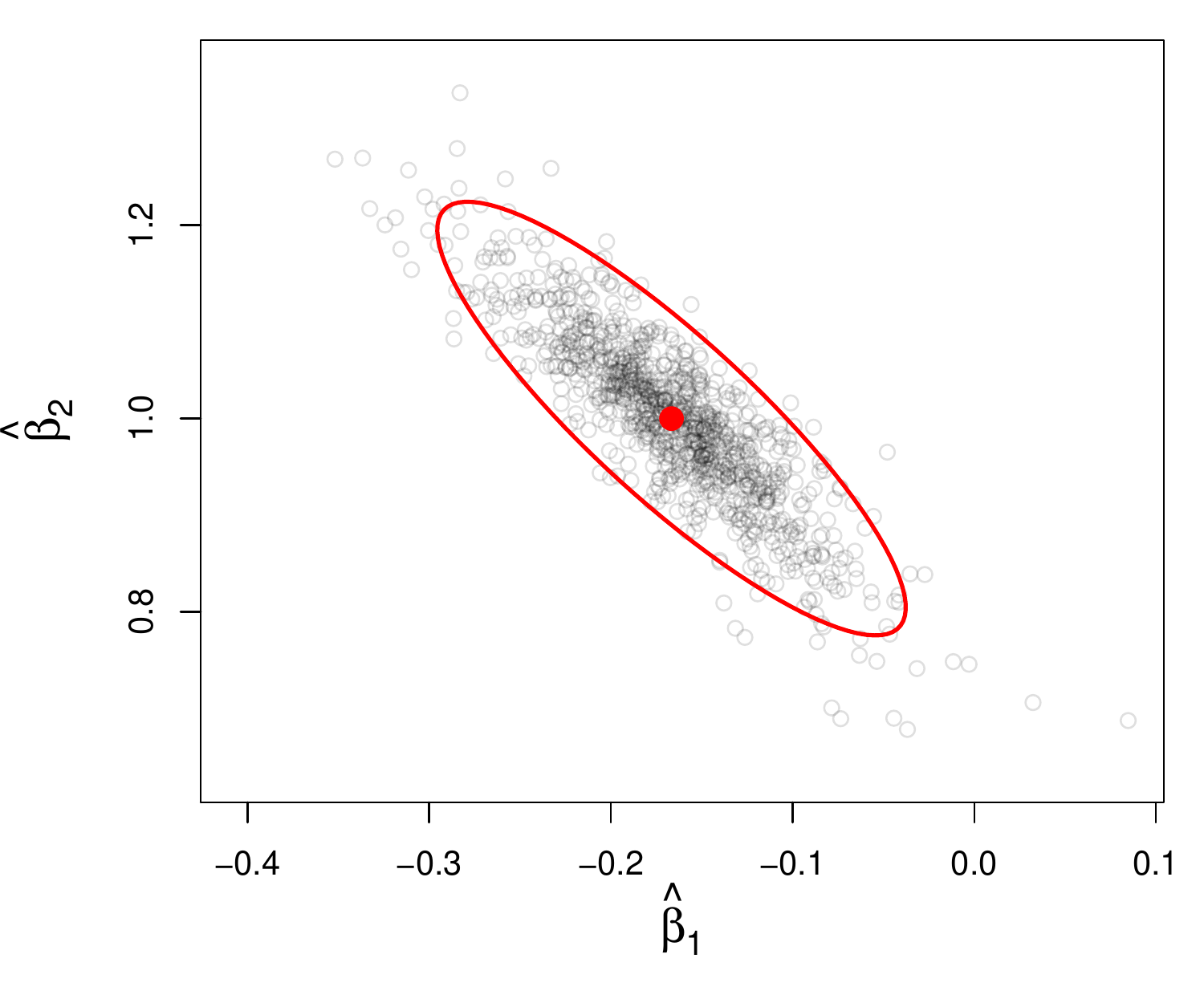}
 } 
\\
 \subfloat[$\widehat{W}=(\Sigma + \widehat{\Delta})^{-1}$]{\label{fig:estimator_dls}
\includegraphics[scale=\w]{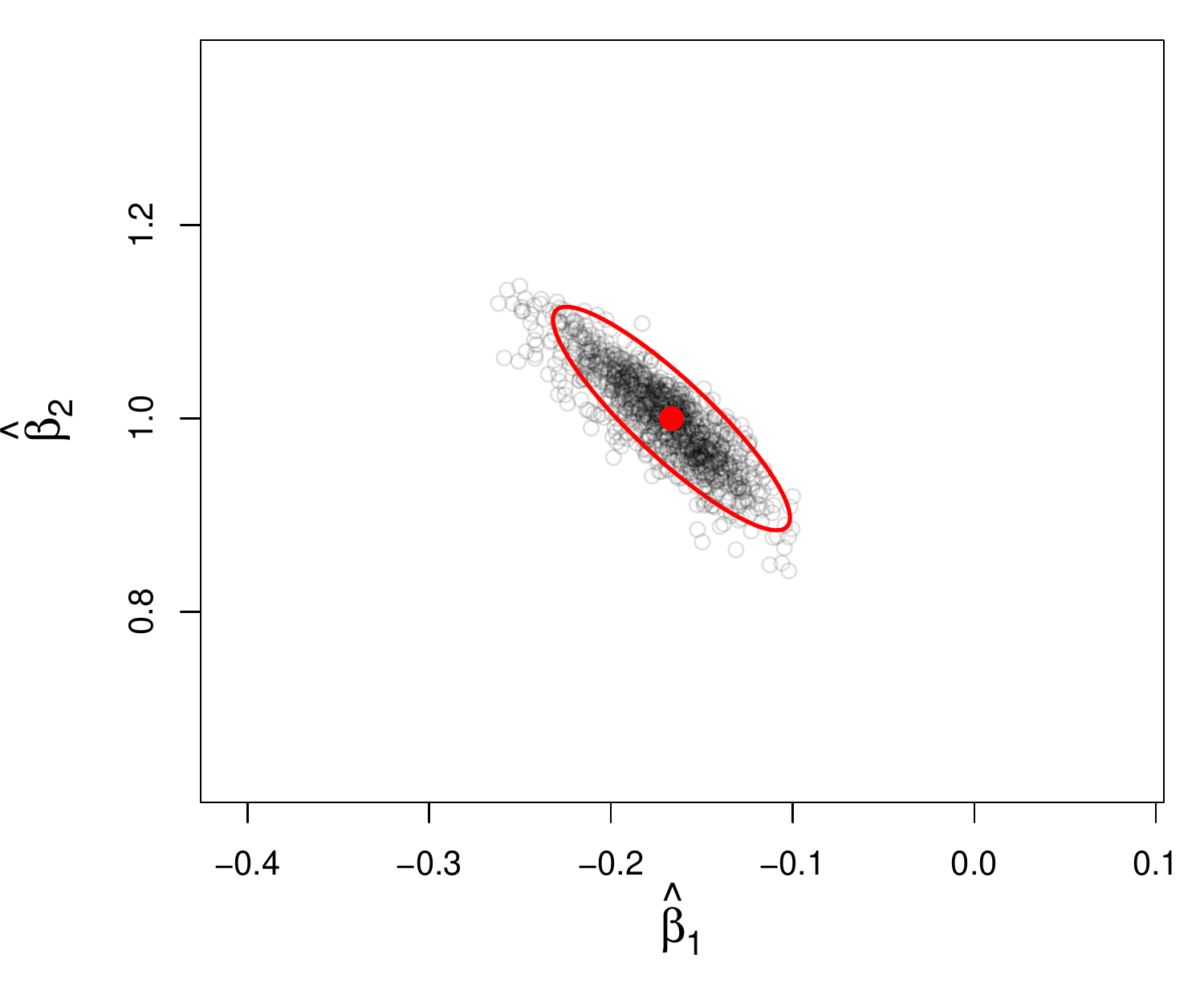}
 } 
 \subfloat[$\widehat{W}_{min,ii} =  \frac{\Gamma(\widehat{B})}{\Gamma(\widehat{B}^T\widehat{C}_{m_i}\widehat{B})}$]{\label{fig:estimator_uls}
\includegraphics[scale=\w]{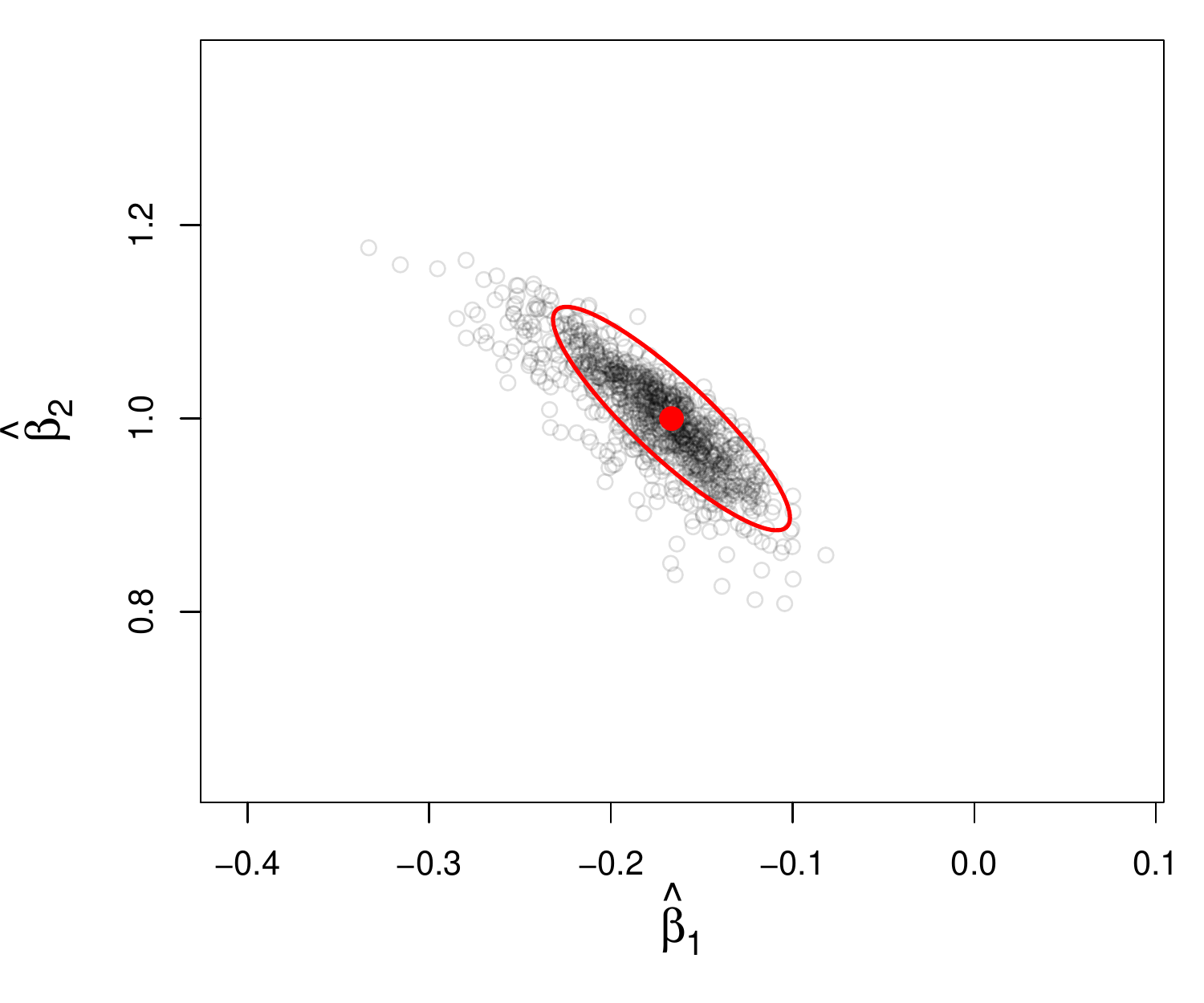}
 } 
 \caption{Parameter estimates using (a) standard WLS (b) OLS (c) estimated weights assuming the $\sigma_i$ are known (d) estimated weights using only the group membership of the variances. The red ellipses are the asymptotic variances of the various methods. \label{fig:estimator_distributions}}
\end{figure}

% latex table generated in R 3.3.2 by xtable 1.8-2 package
% Tue Jan 10 13:58:10 2017
\begin{table}[ht]
\centering
\begin{tabular}{ccccc}
  \hline
 & WLS & OLS & $(\Sigma + \widehat{\Delta})^{-1}$ & $\frac{\Gamma(\widehat{B})}{\Gamma(\widehat{B}^T\widehat{C}_{m_i}\widehat{B})}$ \\ 
  \hline
$\widehat{\nu}_1$ & 0.536 & 0.945 & 0.807 & ----- \\ 
  $\widehat{\nu}_2$ & 0.393 & 0.96 & 0.843 & 0.759 \\ 
  $\widehat{\nu}_{OR}$ & 0.925 & 0.945 & 0.956 & ----- \\ 
   \hline
\end{tabular}
\caption{Fraction of times $\beta$ is in 95\% confidence region.} 
\label{tab:CI}
\end{table}

We conduct a small simulation study to demonstrate some of the ideas presented in the last section.\footnote{Code to reproduce the work in this section can be accessed at \url{http://stat.tamu.edu/~jlong/hetero.zip} or by contacting the author.} Consider modeling the function $f(x) = x^2$ using linear regression with an intercept term. Let $x \sim Unif(0,1)$. The best linear approximation to $f$ is $\beta_1 + \beta_2x$ where $\beta_1 = -1/6$ and $\beta_2=1$. We first suppose $\sigma$ is drawn independently from $x$ from a discrete probability distribution such that $P(\sigma=0.01) = P(\sigma=1) = 0.05$ and $P(\sigma=0.1)=0.9$. Since $\sigma$ has support on a finite set of values, we can consider the cases where $\sigma_i$ is known (Section \ref{sec:known}) and where only the group $m_i$ of observation $i$ is known (Section \ref{sec:unknown}). We let $\Gamma$ be the trace of the matrix.

We generate samples of size $n=100$, $N=1000$ times and make scatterplots of the parameter estimates using weights $W=\Sigma^{-1}$ (standard WLS), $W=I$ (OLS), $\widehat{W}_{min}=(\Sigma + \widehat{\Delta})^{-1}$, and $\widehat{W}_{min,ii} =  \frac{\Gamma(\widehat{B})}{\Gamma(\widehat{B}^T\widehat{C}_{m_i}\widehat{B})}$. The OLS estimator does not require any knowledge about the $\sigma_i$. The fourth estimator uses only the group $m_i$ of observation $i$. For the two adaptive estimators, we use $\widehat{\beta}(I)$ as an initial root $n$ consistent estimator of $\beta$ and iterate twice to obtain the weights.

Results are shown in Figure \ref{fig:estimator_distributions}. The red ellipses are the asymptotic variances. The results show that OLS outperforms standard WLS. Estimating the optimal weighting with or without knowledge of the variances outperforms both OLS and standard WLS. Exact knowledge of the weights (c) somewhat outperforms only knowing the group membership of the variances (d).

We construct 95\% confidence regions using estimates of the asymptotic variance and determine the fraction of times (out of the $N$ simulations) that the true parameters are in the confidence regions. Recall that in Section \ref{sec:known} we proposed $\widehat{\nu}_1$ (Equation \eqref{eq:nuhat1}) as well as the oracle $\widehat{\nu}_{OR}$ for estimating the asymptotic variance when the error variances are known. In Section \ref{sec:unknown} we proposed $\widehat{\nu}_2$ (Equation \eqref{eq:nuhat2}) when the error variances are unknown. The estimator $\widehat{\nu}_2$ can also be used when the error variances are known. We use all three of these methods for constructing confidence regions for standard WLS, OLS, and $\widehat{W} = (\Sigma + \widehat{\Delta})^{-1}$. For $\widehat{W}_{ii} = \frac{\Gamma(\widehat{B})}{\Gamma(\widehat{B}^T\widehat{C}_{m_i}\widehat{B})}$ we use only $\widehat{\nu}_2$ because $\widehat{\nu}_1$ requires knowledge of $\Sigma$. Table \ref{tab:CI} contains the results. While for OLS the nominal coverage probability is approximately attained, the other methods are anti--conservative for $\widehat{\nu}_1$ and $\widehat{\nu}_2$. Estimates for WLS are especially poor. The performance of the oracle is rather good, suggesting that the problem lies in estimating $A$ and $B$. 

\def \w {.35}
  \begin{figure}[H]
\centering
 \subfloat[$\widehat{W}=\Sigma^{-1}$]{\label{fig:estimator_wls_dep}
\includegraphics[scale=\w]{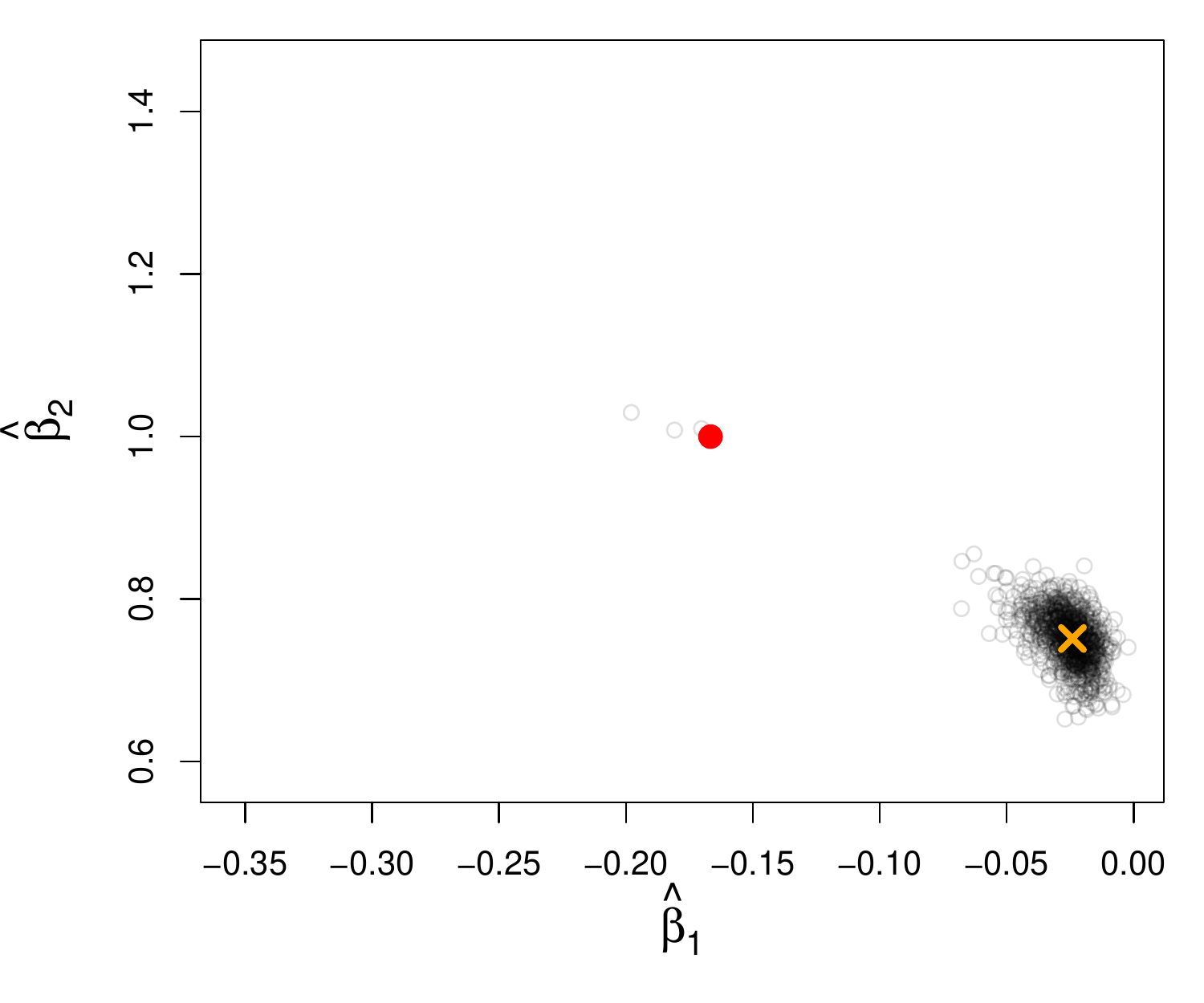}
}
\subfloat[$\widehat{W}=I$]{\label{fig:estimator_ols_dep}
\includegraphics[scale=\w]{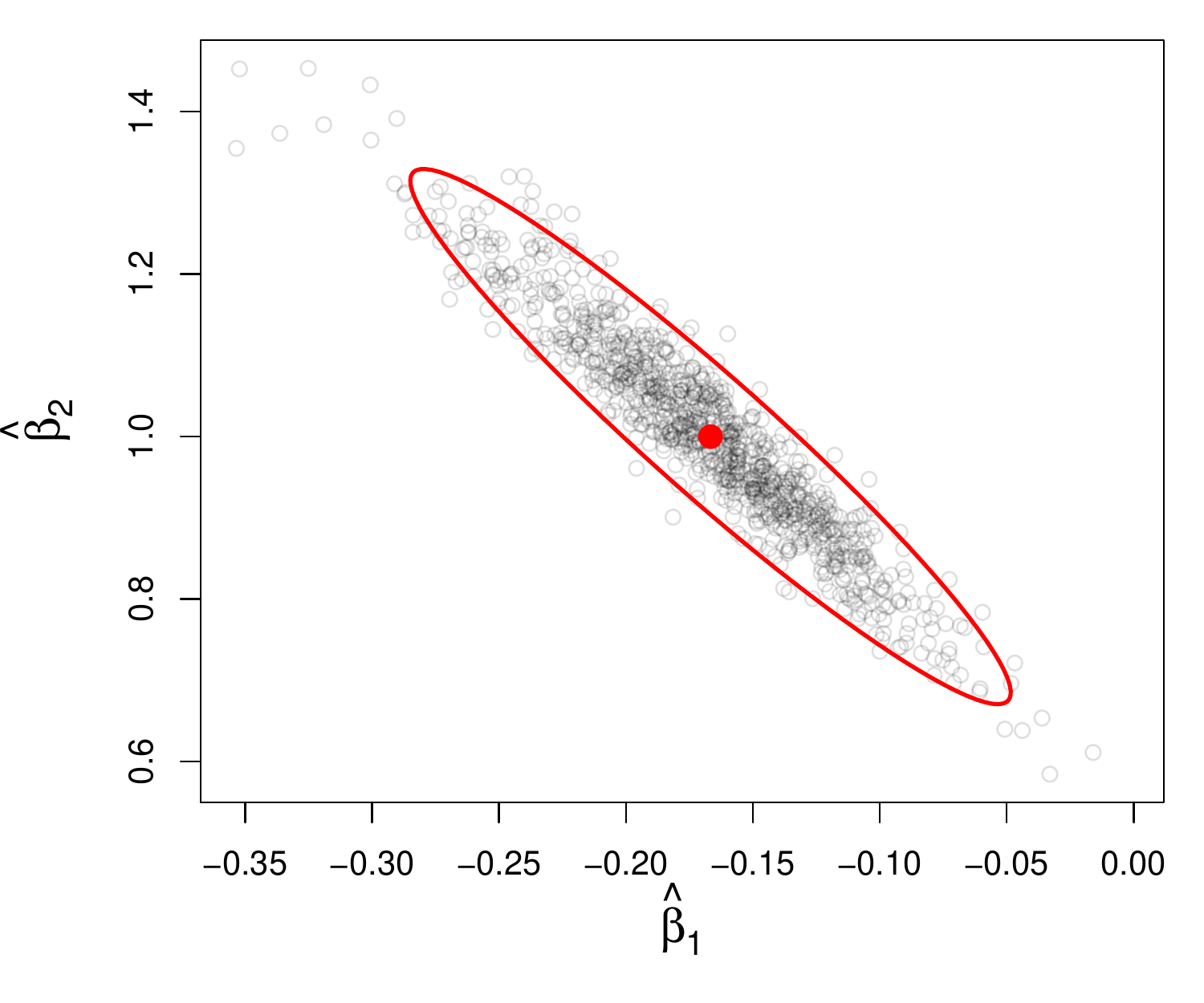}
 } 
 \caption{Parameter estimates using (a) standard WLS and (b) OLS when there is dependence between $x$ and $\sigma$. We see that WLS is no longer consistent. The red point in each plot is the true parameter values. The orange $\times$ in the left plot is the value to which standard WLS is converging (r.h.s. of Equation \eqref{eq:dep_errors}).  The red ellipse is the OLS sandwich asymptotic variance for the dependent case, Equation \eqref{eq:sandwich}. \label{fig:estimator_distributions_dep}}
\end{figure}

To illustrate the importance of the $\sigma$, $x$ independence assumption, we now consider the case where $\sigma$ is a function of $x$. Specifically,
\begin{equation*}
  \sigma =    \left\{
  \begin{array}{lr}
    0.01 &:  x<0.05\\ 
    0.1  &: 0.05\leq x \leq 0.95\\
    1  &: x > 0.95
  \end{array} \right.
\end{equation*}
All other parameters in the simulation are the same as before. Note that the marginal distribution of $\sigma$ is the same as the first simulation. We know from Section \ref{sec:dependent} that weighted estimators may no longer be consistent. In Figure \ref{fig:estimator_distributions_dep} we show a scatter plot of parameter estimates using standard WLS and OLS. We see that the WLS estimator has low variance but is highly biased. The OLS estimator is strongly preferred.

\subsection{Analysis of Astronomy Data}

\cite{sesar2010light} identified 483 RR Lyrae periodic variable stars in Stripe 82 of the Sloan Digital Sky Survey III. We obtained 450 of these light curves from a publicly available data base \cite{ivezic2007sloan}.\footnote{We use only the g--band data for determining periods.} Figure \ref{fig:unfolded} shows one of these light curves. These light curves are well observed ($n > 50$), so it is fairly easy to estimate periods. For example, \cite{sesar2010light} used a method based on the Supersmoother algorithm of \cite{friedman1984variable}. However there is interest in astronomy in developing period estimation algorithms that work well on poorly sampled light curves \cite{vanderplas2015periodograms,mondrik2015multiband,long2014estimating,sesar2007exploring}. Well sampled light curves offer an opportunity to test period estimation algorithms because ground truth is known and they can be artificially downsampled to create realistic simulations of poorly sampled light curves.

\begin{figure}[t]
  \begin{center}
    \begin{includegraphics}[scale=.5]{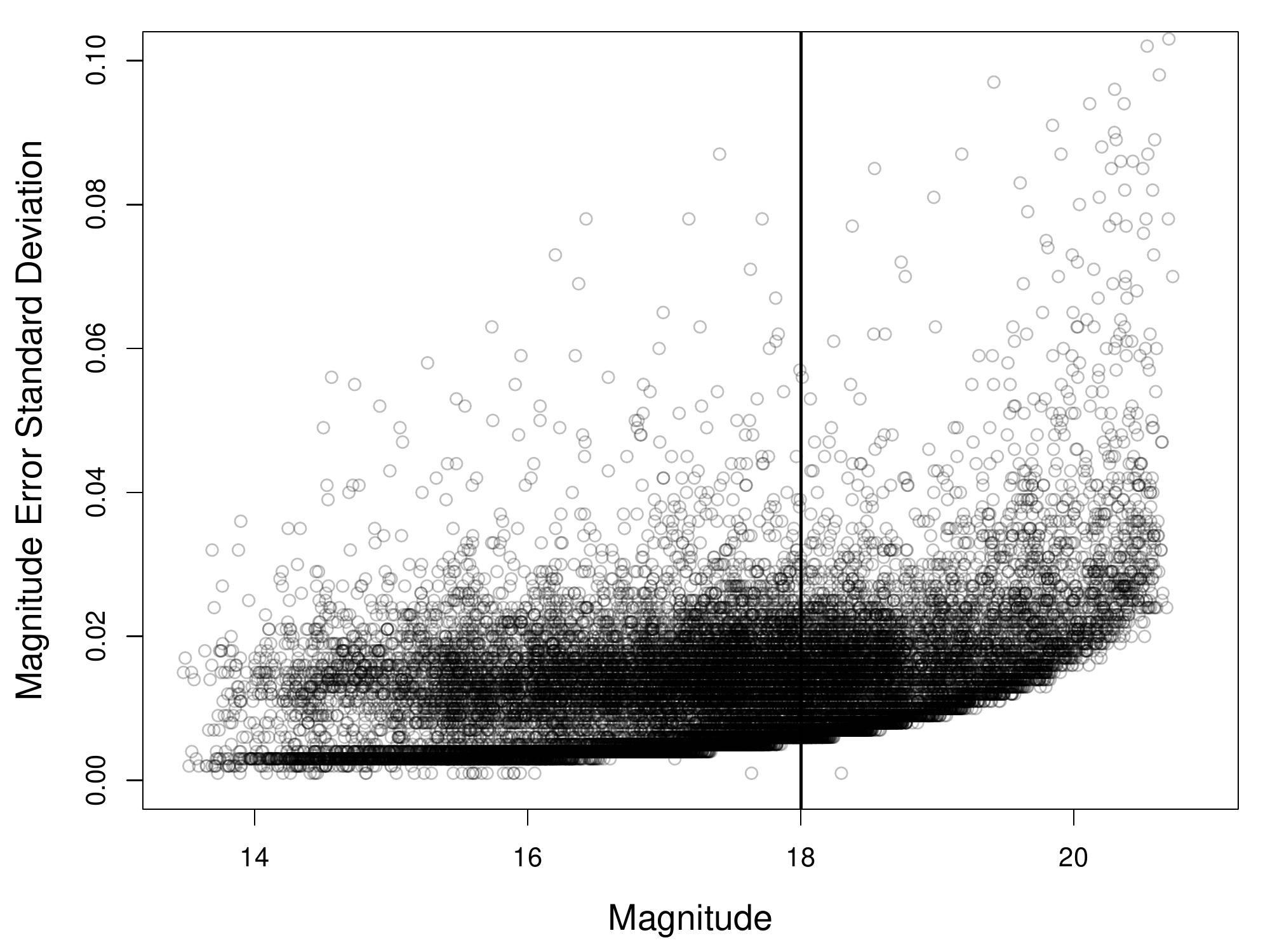}
      \caption{Magnitude error versus magnitude scatterplot. As magnitude increases (observation is less bright), the uncertainty rises. We use only stars where all photometric measurements are less than 18 magnitudes. In this region magnitude error and magnitude are approximately independent.\label{fig:mag_error}}
    \end{includegraphics}
  \end{center}
\end{figure}

As discussed in Section \ref{sec:astro}, each light curve can be represented as $\{(t_i,y_i,\sigma_i)\}_{i=1}^n$ where $t_i$ is the time of the $y_i$ brightness measurement made with uncertainty $\sigma_i$. In Figure \ref{fig:mag_error} we plot magnitude error ($\sigma_i$) against magnitude ($y_i$) for all observations of all 450 light curves. For higher magnitudes (less bright observations), the observation uncertainty is larger.  In an attempt to ensure independence between $\sigma$ and $x$ assumed by our asymptotic theory, we use only the bright stars in which all magnitudes are below 18 (left of the vertical black line in Figure \ref{fig:mag_error}). In this region, magnitude and magnitude error are approximately independent. This reduces the sample to 238 stars. We also ran our methods on the larger set of stars. Qualitatively, the results which follow are similar.

In order to simulate challenging period recovery settings, we downsample each of these light curves to have $n=10,20,30,40$. We estimate periods using sinusoidal models with $K=1,2,3$ harmonics. For each model we consider three methods for incorporating the error variances. In the first two methods, we weight by the the inverse of the observations variances ($\Sigma^{-1}$) as suggested by maximum likelihood for correctly specified models and the identity matrix ($I$). Since this is not a linear model, it is not possible to directly use the weighting idea proposed in Section \ref{sec:known}. We propose a modification for the light curve scenario. We first fit the model using identity weights and determine a best fit period. We then determine the optimal weighting at this period following the procedure of Section \ref{sec:known}. Recall from Section \ref{sec:astro} that at a fixed period, the sinusoidal models are linear. Using the new weights, we then refit the model and estimate the period. A period estimate is considered correct if it is within 1\% of the true value. 

\begin{table}[ht]
\centering
\begin{tabular}{c|ccc|ccc|ccc}
 &  & $K=1$ & &   & $K=2$ &  &  & $K=3$ &  \\ 
  \hline
 & $\Sigma^{-1}$ &  $I$ & $\Delta$ & $\Sigma^{-1}$ &  $I$ & $\Delta$ & $\Sigma^{-1}$ &  $I$ & $\Delta$\\
  \hline10&0.09&0.16&0.15&0.13&0.11&0.11&0.03&0.03&0.03\\20&0.46&0.58&0.59&0.63&0.68&0.69&0.69&0.77&0.77\\30&0.64&0.78&0.79&0.71&0.82&0.83&0.82&0.86&0.85\\40&0.75&0.79&0.79&0.80&0.85&0.85&0.87&0.92&0.92\\\hline
\end{tabular}
\caption{Fraction of periods estimated correctly using different weightings for models with $K=1,2,3$ harmonics. Ignoring the observation uncertainties ($I$) in the fitting is superior to using them ($\Sigma^{-1}$). The strategy for determining an optimal weight function ($\Delta$) does not provide much improvement over ignoring the weights. More complex models ($K=3$) perform worse than simple models ($K=1$) when there is limited data ($n=10$), but better when the functions are better sampled ($n=40$). The standard errors on these accuracies is no larger than $\sqrt{0.5(1-0.5)/238} \approx 0.032$ .}
\label{tab:period_est_results}
\end{table}

The fraction of periods estimated correctly are contained in Table \ref{tab:period_est_results}. In nearly all cases ignoring observation uncertainties ($I$) outperforms using the inverse of the observation variances as weights ($\Sigma^{-1}$). The improvement is greatest for the $K=1$ model and least for the $K=3$ model, possibly due to the decreasing model misspecification as the number of harmonics increases. The very poor performance of the $K=3$ models with 10 magnitude measurements is due to overfitting. With $K=3$, there are 8 parameters which is too complex a model for 10 observations. Optimizing the observation weights does not appear to improve performance over not using weights. This is potentially due to the fact that the model is highly misspecified (see Figure \ref{fig:folded}).

\section{Discussion}
\label{sec:discussion}

\subsection{Other Problems in Astronomy}

Heteroskedastic measurement error is ubiquitous in astronomy problems. In many cases some degree of model misspecification is present. In this work, we focused on the problem of estimating periods of light curves. Other problems include:
\begin{itemize}
\item \cite{salmon2015relation} observe the brightness of galaxies through several photometric filters. Variances on the brightness measurements are heteroskedastic. The brightness measurements for each galaxy are matched to a set of templates. Assuming a normal measurement error model, maximum likelihood would suggest weighting the difference between observed brightness and template brightness by the inverse of the observation variance. In personal communication, \cite{salmon2015relation} stated that galaxy templates contained some level of misspecification. \cite{salmon2015relation} addressed this issue by inflating observation variances, using weights of $(\sigma^2 + \Delta)^{-1}$ instead of $\sigma^{-2}$. The choice of $\Delta > 0$ was based on qualitative analysis of model fits. Section \ref{sec:olswls} provides a theoretical justification for this practice.
\item \cite{richards2012prototype} models spectra of galaxies as linear combinations of simple stellar populations (SSP) and non--linear distortions. While parameters which define an SSP are continuous, a discrete set of SSPs are selected as prototypes and the galaxies are modeled as linear combinations of the prototypes. This is done for computational efficiency and to avoid overfitting. However prototype selection introduces some degree of model misspecification as the prototypes may not be able to perfectly reconstruct all galaxy spectra. Galaxy spectra are observed with heteroskedastic measurement error and the inverse of the observation variances are used as weights when fitting the model (see Equation 2.2 in \cite{richards2012prototype}).
\end{itemize}

\subsection{Conclusions}

We have shown that WLS estimators can perform poorly when the response is not a linear function of the predictors because observations with small variance have too much influence on the fit. In the misspecified model setting, OLS suffers from the usual problem that observations with large variance induce large asymptotic variance in the parameter estimates. For cases in which some observations have very small variance and other observations have very large variance, procedures which optimize the weights may achieve significant performance improvements as shown in the simulation in Section \ref{sec:sim}.

This work primarily focused on the case where $x$ and $\sigma$ are independent. However results from Section \ref{sec:dependent} showed that when independence fails, weighted estimators will typically be biased. This additional complication makes OLS more attractive relative to weighted procedures.

For practitioners we recommend caution in using the inverse of the observation variances as weights when model misspecification is present. As a check, practitioners could fit models twice, with and without weights, and compare performance based on some metric. More sophisticated methods, such as specifically tuning weights for optimal performance may be attempted. Our asymptotic theory provides guidance on how to do this in the case of the linear model.

\appendix

\section{Technical Notes}
\label{sec:technical_notes}

\subsection{Proof of Theorem \ref{thm:clt}}
\label{prf:clt}
Let $g(X) \in \mathbb{R}^n$ be the function $g$ applied to the rows of $X$. We sometimes write $w$ for $w(\sigma)$. We have
\begin{align*}
\widehat{\beta}(\widehat{W}) &= (X^T\widehat{W}X)^{-1}X^T\widehat{W}Y\\
&= (X^T\widehat{W}X)^{-1}X^T\widehat{W}(X\beta + g(X) + \Sigma^{1/2}\epsilon)\\
&= \beta + \underbrace{((1/n)X^T\widehat{W}X)^{-1}}_{\equiv q}\underbrace{(1/n)X^T\widehat{W}(g(X) + \Sigma^{1/2}\epsilon)}_{\equiv z}.
\end{align*}
In part \ref{part:i} we show that 
\begin{equation*}
q \rightarrowp \E[xx^T]^{-1}\E[w]^{-1}.
\end{equation*}
In part \ref{part:ii} we show that
\begin{equation*}
\sqrt{n}z \rightarrowd N(0,\E[w^2]\E[g^2xx^T] + \E[\sigma^2w^2]\E[xx^T]).
\end{equation*}
Thus by Slutsky's Theorem
\begin{align*}
  &\sqrt{n}(\widehat{\beta}(\widehat{W}) - \beta) \\
  &= q\sqrt{n}z \\
&\rightarrowd N\left(0,\E[w]^{-2}(\E[w^2]\E[xx^T]^{-1}\E[g^2(x)xx^T]\E[xx^T]^{-1} + \E[\sigma^2w^2]\E[xx^T]^{-1})\right)
\end{align*}
\begin{enumerate}
\item \label{part:i} \textbf{Show $q \rightarrowp \E[xx^T]^{-1}\E[w]^{-1}$:} 
Recall that by Assumptions \ref{ass:weight}
\begin{equation*}
\widehat{W}_{ii} = w(\sigma_i) + n^{-1/2}\delta_{nm_i}h(\sigma_i) + n^{-1}d(\sigma_i,\delta_n)
\end{equation*}
where $h$ is a bounded function, $\delta_{nm_i}$ are $O_P(1)$, and the $d$ is uniformly (in $\sigma$) bounded by an $O_P(1)$ random variable.
\begin{align*}
q^{-1} &= (1/n)X^T\widehat{W}X\\
&= \frac{1}{n} \sum x_i x_i^T \widehat{W}_{ii}\\
&= \frac{1}{n} \sum x_i x_i^T w(\sigma_i) + \underbrace{ \frac{1}{n^{3/2}}\sum x_i x_i^Th(\sigma_i)\delta_{nm_i}}_{\equiv R_1} + \underbrace{\frac{1}{n^2}\sum x_ix_i^Td(\sigma_i,\delta_n)}_{\equiv R_2}
\end{align*}
We show that $R_1,R_2 \rightarrowp 0$. Noting that $\E[|x_{ij}x_{ik}h(\sigma_i)\ind{m_i=m}|] < \infty$ because $h$ is bounded and the $x$ have second moments we have
\begin{equation*}
|R_{1jk}| = n^{-1/2}\left|\sum_{m=1}^M \delta_{nm}\left(n^{-1}\sum_{i=1}^n x_{ij} x_{ik}h(\sigma_i)\ind{m_i=m}\right)\right| \rightarrowp 0.
\end{equation*}
Using  the fact that $|d(\sigma_i,\delta_n)| < \delta_n'$ where $\delta_n'$ is $O_P(1)$  we have
\begin{equation*}
|R_{2jk}| \leq n^{-1}\delta_n' \left(\frac{1}{n} \sum_{i=1}^n |x_{ij}x_{ik}|\right) \rightarrowp 0.
\end{equation*}

Thus 
\begin{equation*}
  q^{-1} \rightarrowp \E[xx^Tw] = \E[xx^T]\E[w]
\end{equation*}
where the last equality follows from the facts that $\sigma$ and $x$ are independent. The desired result follows from the continuous mapping theorem.
\item \textbf{Show $\sqrt{n}z \rightarrowd N(0,\E[w^2]\E[g^2xx^T] + \E[\sigma^2w^2]\E[xx^T])$:}\label{part:ii} 
\begin{align*}
\sqrt{n}z &= n^{-1/2} \sum_{i=1}^n (g(x_i) + \sigma_i\epsilon_i)\widehat{W}_{ii}x_{i}\\
&=n^{-1/2} \sum_{i=1}^n \underbrace{(g(x_i) + \sigma_i\epsilon_i)w(\sigma_i)x_{i}}_{a_i} + \underbrace{n^{-1}\sum_{i=1}^n (g(x_i) + \sigma_i\epsilon_i)x_i\delta_{nm_i}h(\sigma_i)}_{R_3} \\
&+ \underbrace{n^{-3/2}\sum_{i=1}^n(g(x_i) + \sigma_i\epsilon_i)d(\sigma_i,\delta_n)x_i}_{R_4}
\end{align*}
$\E[a_i] = \E[(g(x_i) + \sigma_i\epsilon_i)w(\sigma_i)x_{i}] = 0$ because $\E[g(x_i)x_i] = 0$ and $\epsilon_i$ is independent of all other terms and mean $0$. We have 
\begin{align*}
Cov(a_i)_{jk} &= \E[a_{ij}a_{ik}] \\
&= \E[(g(x) + \sigma\epsilon)^2w^2x_jx_k]\\
&= \E[g^2(x)w^2x_jx_k] + 2\E[g(x)\sigma\epsilon w^2x_jx_k] + \E[\sigma^2\epsilon^2w^2x_jx_k]\\
&= \E[w^2]\E[g^2(x)x_jx_k] + \E[\sigma^2w^2]\E[x_jx_k].
\end{align*}
So $Cov(a_i) = \E[w^2]\E[g^2xx^T] + \E[\sigma^2w^2]\E[xx^T]$.  The desired result now follows from the CLT and showing that $R_3,R_4 \rightarrowp 0$. Note that
\begin{align*}
  &\E[(g(x_i) + \sigma_i\epsilon_i)x_ih(\sigma_i)\ind{m_i=m}]\\
  &= \E[g(x_i)x_i]\E[h(\sigma_i)\ind{m_i=m}] + \E[\sigma_i\epsilon_ix_ih(\sigma_i)\ind{m_i=m}]\\
  &= 0.
\end{align*}
Thus
\begin{equation*}
R_3 = \sum_{m=1}^M \left(\delta_{nm} n^{-1} \sum_{i=1}^n (g(x_i) + \sigma_i\epsilon_i)x_ih(\sigma_i) \ind{m_i=m}\right) \rightarrowp 0
\end{equation*}
because the terms inside the $i$ summand are i.i.d. with expectation $0$. Finally recalling that the $d(\sigma_i,\delta_n)$ is bounded above by $\delta'_n$ which is uniform $O_P(1)$, we have
\begin{equation*}
|R_{4}| \leq n^{-1/2}\delta'_n\frac{1}{n}\sum_{i=1}^n\left|(g(x_i) + \sigma_i\epsilon_i)x_{i}\right| \rightarrowp 0.
\end{equation*}

\end{enumerate}

\subsection{Proof of Theorem \ref{thm:lagrange}}
\label{prf:lagrange}

Since $w > 0$, by Cauchy Schwartz
\begin{equation*}
\Gamma(\nu(w)) = \frac{\E[w^2(\Gamma(A) + \sigma^2\Gamma(B))]}{\E[w]^2} \geq \E[(\Gamma(A) + \sigma^2\Gamma(B))^{-1}]^{-1}
\end{equation*}
with equality iff
\begin{equation*}
  w(\sigma) \propto \frac{1}{\Gamma(A) + \sigma^2\Gamma(B)} \propto (\sigma^2 + \Gamma(A)\Gamma(B)^{-1})^{-1}
\end{equation*}
with probability 1.

\subsection{Proof of Corollary \ref{thm:adaptive}}
\label{prf:adaptive}
We must show
\begin{equation*}
\Gamma(\nu(w_{min})) \leq \min(\Gamma(\nu(I)),\Gamma(\nu(\Sigma^{-1})))
\end{equation*}
with strict inequality if $\E[g^2(x)xx^T]$ is positive definite and the distribution of $\sigma$ is not a point mass. The inequality follows from Theorem \ref{thm:lagrange}. By Theorem \ref{thm:lagrange}, the inequality is strict whenever the functions $w(\sigma) = 1$ and $w(\sigma) = \sigma^{-2}$ are not proportional to $w_{min}(\sigma) = (\sigma^2 + \Gamma(A)\Gamma(B)^{-1})^{-1}$ with probability $1$. Since $B \succ 0$ and $A = B^T\E[xx^Tg(x)^2]B \succ 0$, $\Gamma(A)\Gamma(B)^{-1} > 0$. So if $\sigma$ is not constant with probability $1$, $P(w_{min}(\sigma) = c) < 1$ for any $c$. Therefore $w_{min}$ is not proportional to $w(\sigma) = 1$ with probability $1$. Similarly, for $w_{min}$ to be proportional to $w(\sigma) = \sigma^{-2}$, there must exist a $c$ such that
\begin{equation*}
1 = P(\sigma^2 + \Gamma(A)\Gamma(B)^{-1} = c\sigma^2) = P(\Gamma(A)\Gamma(B)^{-1} = \sigma^2(c-1)).
\end{equation*}
However since the constant $\Gamma(A)\Gamma(B)^{-1} > 0$ and $\sigma$ is not a point mass, such a $c$ does not exist.

\subsection{Proof of Theorem \ref{thm:satisfy}}
\label{prf:satisfy}

Let $\Delta = \Gamma(A)\Gamma(B)^{-1}$. In part \ref{part:i_sat} we show that
\begin{equation*}
\widehat{\Delta} = \Delta + n^{-1/2}\delta_n
\end{equation*}
where $\delta_n$ is $O_P(1)$. In part \ref{part:ii_sat} we show that
\begin{equation*}
\frac{1}{\sigma_i^2 + \widehat{\Delta}} = \underbrace{\frac{1}{\sigma_i^2 + \Delta}}_{\equiv w(\sigma_i)} + n^{-1/2} \delta_n h(\sigma_i) + n^{-1}d(\sigma_i,\delta_n)
\end{equation*}
where $\delta_n$ is $O_P(1)$, $d(\sigma_i,\delta_n)$ is bounded uniformly by an $O_P(1)$ random variable, and $h$ is a bounded function. Thus the weight matrix $\widehat{W}$ with diagonal elements $\widehat{W}_{ii} = (\sigma_i^2 + \widehat{\Delta})^{-1}$ satisfies Assumptions \ref{ass:weight} with $w(\sigma) = w_{min}(\sigma)$.
\begin{enumerate}
\item \label{part:i_sat} Recall $B = \E[xx^T]^{-1}$. Let $\delta_n$ be $O_P(1)$ which changes definition at each appearance. Define $\widehat{B}^{-1} = n^{-1}X^TX$. By the delta method we have
\begin{equation}
\label{eq:b}
\widehat{B} = B + n^{-1/2}\delta_n
\end{equation}
and
\begin{equation}
\label{eq:basymp}
\Gamma(\widehat{B}) = \Gamma(B) + n^{-1/2}\delta_n.
\end{equation}
By assumption $\widehat{\beta}(\widehat{W}) = \beta + n^{-1/2}\delta_n$, thus
\begin{align*}
  &\left(\sum \sigma_i^{-4}\right)^{-1} \sum \sigma_i^{-4} x_ix_i^T \widehat{g}(x_i)^2 \\
  &= \left(\sum \sigma_i^{-4}\right)^{-1} \sum \sigma_i^{-4} x_ix_i^T ((y_i - x_i^T\widehat{\beta}(\widehat{W}))^2 - \sigma_i^2)\\
&= \left(\sum \sigma_i^{-4}\right)^{-1} \sum \sigma_i^{-4} x_ix_i^T ((y_i - x_i^T\beta)^2 - \sigma_i^2) + n^{-1/2}\delta_n\\
&= \frac{\E[\sigma^{-4}]}{n^{-1}\sum \sigma_i^{-4}}\frac{1}{n} \sum \E[\sigma^{-4}]^{-1}\sigma_i^{-4} x_ix_i^T ((y_i - x_i^T\beta)^2 - \sigma_i^2) + n^{-1/2}\delta_n
\end{align*}
Note that $\E[\sigma^{-4}](n^{-1}\sum \sigma_i^{-4})^{-1} \rightarrowp 1$. Further note that $\E[\sigma^{-4}]^{-1}\sigma_i^{-4} x_ix_i^T ((y_i - x_i^T\beta)^2 - \sigma_i^2)$ are i.i.d. with expectation $\E[xx^Tg(x)^2]$. Thus by the CLT and Slutsky's Theorem
\begin{equation}
\label{eq:ap}
\left(\sum \sigma_i^{-4}\right)^{-1}\sum \sigma_i^{-4}x_ix_i^T\widehat{g}(x_i)^2 = \E[xx^Tg(x)^2] + n^{-1/2}\delta_n.
\end{equation}
Since $\widehat{A} = \widehat{B}^T\left(\sum \sigma_i^{-4}\right)^{-1}\left(\sum \sigma_i^{-4}x_ix_i^T\widehat{g}(x_i)^2\right)\widehat{B}$, by Equations \eqref{eq:b} and \eqref{eq:ap} we have
\begin{equation*}
\widehat{A} = A + n^{-1/2}\delta_n.
\end{equation*}
which implies
\begin{equation*}
\Gamma(\widehat{A}) = \Gamma(A) + n^{-1/2}\delta_n.
\end{equation*}
Combining this result with Equation \eqref{eq:basymp} we have
\begin{equation*}
\Gamma(\widehat{A})\Gamma(\widehat{B})^{-1} = \underbrace{\Gamma(A)\Gamma(B)^{-1}}_{\equiv \Delta} + \, n^{-1/2}\delta_n.
\end{equation*}
Since $A$ and $B$ are p.s.d., $\Delta \geq 0$. Therefore
\begin{equation*}
|\Delta - \underbrace{\max(\Gamma(\widehat{A})\Gamma(\widehat{B})^{-1},0)}_{\equiv \widehat{\Delta}}| \leq |\Delta - \Gamma(\widehat{A})\Gamma(\widehat{B})^{-1}|.
\end{equation*}
Thus
\begin{equation*}
\widehat{\Delta} = \Delta + n^{-1/2}\delta_n.
\end{equation*}

\item \label{part:ii_sat} From part \ref{part:i_sat}, using the fact that $(1-x)^{-1} = 1 + x + x^2(1-x)^{-1}$, we have
\begin{align*}
\frac{1}{\sigma_i^2 + \widehat{\Delta}} &= \frac{1}{\sigma_i^2 + \Delta + n^{-1/2}\delta_n}\\
&= \left(\frac{1}{\sigma_i^2 + \Delta}\right)\left(\frac{1}{1-\left(\frac{-n^{-1/2}\delta_n}{\sigma_i^2 + \Delta}\right)}\right)\\
&= \left(\frac{1}{\sigma_i^2 + \Delta}\right)\left(1 - \frac{n^{-1/2}\delta_n}{\sigma_i^2 + \Delta} + \frac{\frac{n^{-1}\delta_n^2}{(\sigma_i^2 + \Delta)^2}}{1+ \frac{n^{-1/2}\delta_n}{\sigma_i^2 + \Delta}}\right)\\
&= \frac{1}{\sigma_i^2 + \Delta} - n^{-1/2}\delta_n\underbrace{\frac{1}{(\sigma_i^2 + \Delta)^2}}_{\equiv h(\sigma_i)} + n^{-1}\underbrace{\frac{\delta_n^2(\sigma_i^2 + \Delta)^{-2}}{(\sigma_i^2 + \Delta) + n^{-1/2}\delta_n}}_{\equiv d(\sigma_i,\delta_n)}.
\end{align*}
The function $h$ is bounded because the $\sigma_i$ are bounded below by a positive constant and $\Delta \geq 0$. Note that since $\sigma_i \geq \sigma_{min} > 0$ we have
\begin{equation*}
d(\sigma_i,\delta_n) \leq \frac{\frac{\delta_n^2}{\sigma_{min}^4}}{\sigma_{min}^2 + n^{-1/2}\delta_n}
\end{equation*}
where the right hand side is $O_P(1)$.
\end{enumerate}

\subsection{Proof of Theorem \ref{thm:satisfy2}}
\label{prf:satisfy2}

Let $\delta_n, \delta_{nm}$ be $O_P(1)$ which change definition at each appearance. From Equations \eqref{eq:b} and \eqref{eq:basymp} in Proof \ref{prf:satisfy} we have
\begin{align*}
\widehat{B} &= B + n^{-1/2}\delta_n\\
\Gamma(\widehat{B}) &= \Gamma(B) + n^{-1/2}\delta_n.
\end{align*}
We have
\begin{align*}
\widehat{C}_m &= \frac{1}{\sum_{i=1}^n \ind{m_i=m}} \sum_{i=1}^n (y_i - x_i^T\widehat{\beta}(\widehat{W}))^2x_ix_i^T\ind{m_i=m}\\
&= \frac{nf_m(m)}{\sum_{i=1}^n \ind{m_i=m}} \left(\frac{1}{n} \sum_{i=1}^n \frac{(y_i - x_i^T\beta)^2x_ix_i^T\ind{m_i=m}}{f_m(m)}\right) + n^{-1/2}\delta_{nm}\\
&=C_m + n^{-1/2}\delta_{nm}
\end{align*}
where the last equality follow from the facts that the terms inside the sum are i.i.d with expectation $C_m = \E[(g^2(x) + \sigma_m^2)xx^T]$ and $\frac{nf_m(m)}{\sum_{i=1}^n \ind{m_i=m}} \rightarrow_P 1$. Thus we have
\begin{equation*}
\widehat{W}_{min,ii} = w_{min}(m_i) + \delta_{nm_i}n^{-1/2}
\end{equation*}
which satisfies the form of Assumptions \ref{ass:weight}.

\subsection{Proof of Theorem \ref{thm:dep}}
\label{prf:dep}

\begin{equation*}
\widehat{\beta}(W) = (X^TWX)^{-1}X^TWY = \left(\frac{1}{n}\sum x_ix_i^Tw(\sigma_i)\right)^{-1}\left(\frac{1}{n} \sum x_iw(\sigma_i)y_i\right)
\end{equation*}
By the SLLN and the continuous mapping theorem
\begin{equation*}
\left(\frac{1}{n}\sum x_ix_i^Tw(\sigma_i)\right)^{-1} \rightarrow_{as} \E[xx^Tw(\sigma)]^{-1}.
\end{equation*}
Note that
\begin{equation*}
\frac{1}{n}\sum x_iw(\sigma_i)y_i = \frac{1}{n}\sum x_iw(\sigma_i)f(x_i) + \frac{1}{n}\sum x_iw(\sigma_i)\epsilon_i\sigma_i.
\end{equation*}
The summands in second term on the r.h.s. are i.i.d. with expectation $0$. Therefore
\begin{equation*}
\frac{1}{n}\sum x_iw(\sigma_i)y_i \rightarrow_{as} \E[xw(\sigma)f(x)].
\end{equation*}

\bibliographystyle{abbrvnat}
\bibliography{refs}

%\clearpage\pagebreak\newpage
%% \pagestyle{fancy}
%% \fancyhf{}
%% \rhead{\bfseries\thepage}
%% \lhead{\bfseries Appendix}
%% \begin{center}
%% {\LARGE{\bf Appendix to\\ {\it A Note on Parameter Estimation for Misspecified Regression Models with Heteroskedastic Errors}}}
%% \end{center}

%% \setcounter{equation}{0}
%% \setcounter{page}{1}
%% \setcounter{table}{1}
%% \setcounter{section}{0}
%% \renewcommand{\theequation}{A.\arabic{equation}}
%% \renewcommand{\thesection}{A.\arabic{section}}
%% \renewcommand{\thesubsection}{A.\arabic{section}.\arabic{subsection}}
%% \renewcommand{\thepage}{A.\arabic{page}}
%% \renewcommand{\thetable}{A.\arabic{table}}
%% \baselineskip=17pt

\end{document}